\documentclass[letterpaper,twocolumn,10pt]{article}
\usepackage{usenix-2020-09}

\usepackage{tikz}
\usepackage{amsmath}

\usepackage{amsfonts}
\usepackage{amssymb}
\usepackage{array}
\usepackage{graphicx}
\usepackage{textcomp}
\usepackage{xcolor}
\usepackage{pifont}
\usepackage{multirow}
\usepackage{xspace}
\usepackage{tcolorbox}
\usepackage{wrapfig}
\usepackage{booktabs}
\usepackage{soul}
\usepackage{enumitem}
\usepackage{diagbox}
\usepackage{circledtext}
\circledtextset{resize=real}
\usepackage[linesnumbered,ruled,vlined]{algorithm2e}
\usepackage[labelformat=simple,font=footnotesize]{subcaption}

\usepackage[firstpageonly]{draftwatermark}
\SetWatermarkAngle{0} 
\SetWatermarkLightness{0}
\SetWatermarkFontSize{15pt}
\definecolor{myred}{RGB}{128, 0, 0}
\SetWatermarkColor{myred}
\SetWatermarkVerCenter{0.05\paperwidth}
\SetWatermarkText{\textsf{\bfseries{This paper has been accepted by 33rd USENIX Security Symposium 2024.}}}

\usepackage{url}

\newcommand{\ie}{\hbox{\emph{i.e.}}\xspace}
\newcommand{\eg}{\hbox{\emph{e.g.}}\xspace}
\newcommand{\etc}{\hbox{\emph{etc}}\xspace}
\newcommand{\etal}{\hbox{\emph{et al.}}\xspace}

\newcommand{\tabincell}[2]{\begin{tabular}{@{}#1@{}}#2\end{tabular}}

\DeclareMathOperator*{\argmax}{arg\,max}
\DeclareMathOperator*{\argmin}{arg\,min}
\def\Figref#1{Fig.~\ref{#1}}
\def\Secref#1{$\S\,$\ref{#1}}
\def\Eqref#1{Eq.~(\ref{#1})}

\newcommand{\mytoolsc}{\textsc{MalGuise}\xspace}

\newcommand{\mytool}{\mbox{$\mathsf{MalGuise}$}\xspace}
\newcommand{\mytoolNOSpace}{\mbox{\textsf{MalGuise}}}
\newcommand{\call}{\texttt{call}-\texttt{based} \texttt{redividing}}
\newcommand{\cfg}{{CFG}}
\newcommand{\nop}{{$\mathtt{nop}$}}
\newcommand{\snops}{\text{s-}\mathtt{nop}\text{s}}
\newcommand{\hash}[1]{\textcolor{blue}{\textsf{\zz#1\zz}}}
\def\zz#1{%
 \ifx\zz#1\else
   #1\linebreak[1]\expandafter\zz
 \fi}

\begin{document}
\pagestyle{empty} 
\date{}

\title{\Large \bf A Wolf in Sheep's Clothing: {Practical} {Black-box} Adversarial Attacks for\\{Evading} {Learning}-based Windows Malware Detection in the Wild}

\author{
{\rm
Xiang Ling\textsuperscript{1,2,3}\thanks{Xiang Ling and Zhiyu Wu are the co-first authors. Bin Wang and Jingzheng Wu are the co-corresponding authors.}, ~Zhiyu Wu\textsuperscript{6}, ~Bin Wang\textsuperscript{4,5}, ~Wei Deng\textsuperscript{6}, ~Jingzheng Wu\textsuperscript{1,2,3}}\\
{\rm Shouling Ji\textsuperscript{6}, ~Tianyue Luo\textsuperscript{1}, ~and ~Yanjun Wu\textsuperscript{1,2,3}}\\
{\rm \textsuperscript{1}Intelligent Software Research Center, Institute of Software, Chinese Academy of Sciences}\\
{\rm \textsuperscript{2}Key Laboratory of System Software (Chinese Academy of Sciences)}\\
{\rm \textsuperscript{3}State Key Laboratory of Computer Science, Institute of Software, Chinese Academy of Sciences}\\
{\rm \textsuperscript{4}Zhejiang Key Laboratory of Artificial Intelligence of Things (AIoT) Network and Data Security}\\
{\rm \textsuperscript{5}Hangzhou Research Institute, Xidian University}~~~~{\rm \textsuperscript{6}Zhejiang University}
}

\maketitle

\begin{abstract}
Given the remarkable achievements of existing learning-based malware detection in both academia and industry, this paper presents {\mytool}, a practical black-box adversarial attack framework that evaluates the security risks of existing learning-based Windows malware detection systems under the black-box setting.
{\mytool} first employs a novel semantics-preserving transformation of {\call} to concurrently manipulate both nodes and edges of malware's control-flow graph, making it less noticeable.
By employing a Monte-Carlo-tree-search-based optimization, {\mytool} then searches for an optimized sequence of {\call} transformations to apply to the input Windows malware for evasions.
Finally, it reconstructs the adversarial malware file based on the optimized transformation sequence while adhering to Windows executable format constraints, thereby maintaining the same semantics as the original.
{\mytool} is systematically evaluated against three state-of-the-art learning-based Windows malware detection systems under the black-box setting.
Evaluation results demonstrate that {\mytool} achieves a remarkably high attack success rate, mostly exceeding 95\%, with over 91\% of the generated adversarial malware files maintaining the same semantics.
Furthermore, {\mytool} achieves up to a 74.97\% attack success rate against five anti-virus products, highlighting potential tangible security concerns to real-world users.
\end{abstract}

\section{Introduction}
With the sustainable development of computer technology,  the proliferation of malware, short for \textbf{mal}icious soft\textbf{ware}, has emerged as a grave security threat that performs malicious activities on computer systems.
In particular, the widespread adoption of the Microsoft Windows family of operating systems (\ie, Windows) has rendered it a primary target for malware attacks.
According to AV-TEST~\cite{av_test_about}, the first three quarters of 2022 have witnessed approximately 59.58 million new instances of Windows malware, constituting over 95\% of all recently identified malware samples during this period~\cite{atlas_vpn_news}.
To defend against the ever-increasing Windows malware, considerable research efforts with cutting-edge technologies have been devoted to Windows malware detection~\cite{ye2017survey,ceschin2020machine,ling2023survey}.
Basically, Windows malware detection can trace its history back to signature-based malware detection in the 1990s, which mainly blacklists suspicious software based on a frequently updated database of previously collected malware signatures.
Evidently, signature-based malware detection cannot detect new or previously unknown malware.
In the past two decades, a variety of machine learning (ML) and deep learning (DL) models have been explored and employed for Windows malware detection, collectively referred to as \textit{learning-based Windows malware detection}~\cite{ye2017survey,ling2023survey} in this paper.
Leveraging the high learning capacities of ML/DL models, learning-based Windows malware detection has demonstrated its ability to detect newly emerging and even zero-day malware, establishing itself as a pivotal component of contemporary mainstream anti-virus products in a fiercely competitive market~\cite{kaspersky_AI_powered_antivirus,2022_AI_powered_antivirus}.

It is well known that signature-based malware detection can be easily evaded by traditional obfuscation techniques, such as compression, encryption, register reassignment, code virtualization, \etc.
However, with the widespread availability of de-obfuscation tools and the advanced capabilities of learning-based malware detection, traditional obfuscations are largely ineffective in evading learning-based malware detection~\cite{raff2017malware,kim2022obfuscated,aghakhani2020malware,zhang2022semantics}.
On the other hand, recent studies have unveiled that ML/DL models are inherently vulnerable to adversarial attacks~\cite{goodfellow2014explaining,xu2020adversarial,ling2019deepsec,li2021towards,li2023difficulty} in various domains (\eg, computer vision, natural language processing), by which the adversary maliciously creates adversarial examples as the input to trigger the target ML/DL model to make mistakes.
Therefore, in addition to traditional obfuscations, this paper explores adversarial attacks with a specific focus on evading present-day learning-based Windows malware detection systems.
It should be noted that this paper does not aim to substitute traditional obfuscations, but rather to complement their limitations in evading advanced learning-based malware detection due to its widespread adoption and escalating prominence in both academia and industry~\cite{ye2017survey,kaspersky_AI_powered_antivirus,2022_AI_powered_antivirus}.

In particular, this paper attempts to explore an adversarial attack under the realistic black-box setting for effectively and efficiently generating practical adversarial malware files, which are capable of evading learning-based Windows malware detection systems.
Towards this, we identify two key challenges (\textbf{C\#1} \& \textbf{C\#2}) that need to be addressed as follows.

$\bullet$ \textbf{C\#1}:
\ul{How to generate practical adversarial malware files that maintain the same semantics as the original ones while remaining less noticeable to possible defenders?}
Previous adversarial attacks mainly consist of \circledtext[height=1.5ex]{i} adding irrelevant API calls~\cite{hu2017generating,chen2017adversarial,al2018adversarial,verwer2020robust}, \circledtext[height=1.5ex]{ii} manipulating raw bytes of malware partially or globally~\cite{anderson2017evading,kreuk2018deceiving,suciu2019exploring,kolosnjaji2018adversarial,demetrio2020adversarial,lucas2021malware}, and \circledtext[height=1.5ex]{iii} manipulating the control-flow graph (CFG) of malware by injecting semantic {\nop}s~\cite{zhang2022semantics}.
We argue that the first two types of adversarial attacks (\circledtext[height=1.5ex]{i} \& \circledtext[height=1.5ex]{ii}) are either impractical as they generate adversarial features rather than files, or are strictly limited to specific malware detection like MalConv~\cite{raff2017malware}.
While the third type \circledtext[height=1.5ex]{iii} exhibits improved scalability against {\cfg}-based malware detection, it only considers coarse-grained transformations that manipulate {\cfg}'s nodes, rendering it quite noticeable and easily detectable by defenders.
Thus, to tackle \textbf{C\#1}, we propose a novel fine-grained transformation towards {\cfg}, namely {\call}, which not only manipulates the nodes (\ie, instruction blocks) but also the edges, \ie, the control-flow relationships between two blocks.

$\bullet$ \textbf{C\#2}: 
\ul{How to efficiently search in the vast and discrete space of malware under the black-box setting such that the optimized adversarial malware file can evade learning-based Windows malware detection?}
We investigate existing state-of-the-art black-box adversarial attacks, including gradient estimations with surrogate models~\cite{al2018adversarial,verwer2020robust,kolosnjaji2018adversarial}, evolutionary algorithms~\cite{chen2017adversarial,lucas2021malware}, and reinforcement learning~\cite{anderson2017evading,zhang2022semantics}.
It is evident that those attacks based on gradient estimations heavily rely on prior information (\eg, training data, model architecture) about the target system.
Likewise, those attacks based on evolutionary algorithms and reinforcement learning are computationally expensive due to the vast and discrete space of malware.
Thus, we address \textbf{C\#2} by employing a Monte Carlo tree search (MCTS)-based optimization to efficiently search for the optimal adversarial malware that can successfully evade the target malware detection system.

To sum up, this paper proposes a practical adversarial attack framework, namely {\mytool}, against learning-based Windows malware detection systems under the black-box setting.
As depicted in \Figref{figure:attack_framework}, {\mytool} first represents the input Windows malware as the {\cfg} and introduces a novel semantics-preserving transformation of {\call} that can manipulate both nodes and edges of {\cfg}, making it less noticeable compared to previous attacks.
Then, we employ an MCTS optimization that could effectively and efficiently guide \mytool to search for an optimal sequence of {\call} transformations within the vast and discrete space of malware under the black-box setting.
Finally, based on the optimized transformation sequence, {\mytool} reconstructs the adversarial malware file while adhering to the constraints of Windows executables, thereby successfully evading the target Windows malware detection system while preserving the same semantics as the original malware.

We evaluate the attack effectiveness of \mytool against three representative learning-based Windows malware detection systems (\ie, MalGraph, Magic, and MalConv) compared with two state-of-the-art adversarial attacks on a large wild dataset containing hundreds of thousands of Windows malware and goodware samples.
Evaluation results show that {\mytool} is agnostic to the target learning-based Windows malware detection systems and consistently achieves a high attack success rate exceeding 95\%.
Meanwhile, we empirically verify that {\mytool} can generate realistic adversarial malware files with a probability of over 91\%, a significant improvement over previous adversarial attacks that achieved probabilities of less than 50\% or failed completely.
Furthermore, to understand the security risks of anti-virus products in the wild, we evaluate and observe that the attack success rate of \mytool against five commonly used commercial anti-virus products can reach a range of 11.29\% to 74.97\%.
To summarize, we highlight our key contributions as follows:
\begin{itemize}[leftmargin=10pt,labelwidth=6pt,labelsep=4pt,itemindent=2pt,itemsep=0pt,topsep=0pt,parsep=0pt]
    \item To understand and assess the security risks of present-day learning-based Windows malware detection, we propose a practical black-box adversarial attack framework {\mytool} that generates realistic adversarial malware files.
    
    \item To the best of our knowledge, \mytool is the first to apply a more fine-grained transformation to the {\cfg} of Windows malware, \call, which not only manipulates its nodes (\ie, instruction blocks) but also its edges, \ie, the control-flow relationship.
    
    \item Evaluations demonstrate that \mytool not only effectively evades state-of-the-art learning-based Windows malware detection with attack success rates exceeding 95\%, but also evades five commercial anti-virus products, achieving attack success rates ranging from 11.29\% to 74.97\%.
\end{itemize}
\section{Preliminaries \& Threat Model}\label{sec:pre:threat:model}

\subsection{Preliminaries on learning-based malware detection}\label{subsec:preliminaries:learningmalwaredetection}
\begin{figure}[htbp]
    \centering
    \includegraphics[width=0.90\columnwidth,keepaspectratio]{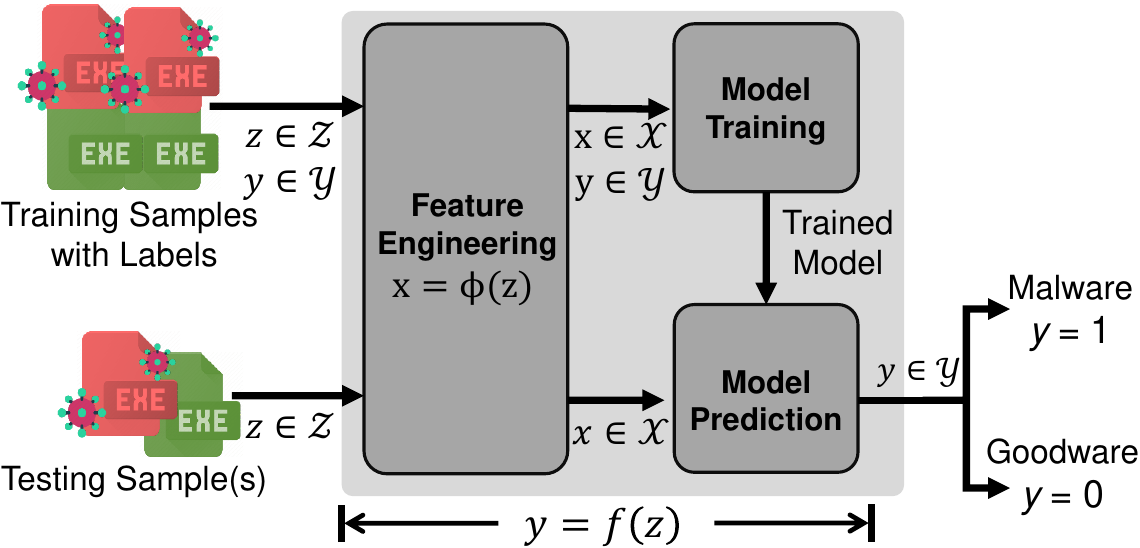}
    \caption{A overview of learning-based malware detection.}
    \label{figure:general_framework}
\end{figure}

\Figref{figure:general_framework} provides an overview of learning-based malware detection.
First, as ML/DL models only operate on numeric data, the training and testing samples are pre-processed by feature engineering before inputting.
Formally, feature engineering can be formulated as $\phi\!:\mathcal{Z}\!\to\!\mathcal{X}$, which produces a feature vector $x$ in the \textit{feature-space} $\mathcal{X}$ (\ie, $x\!\in\!\mathcal{X}$) for a given executable $z$ in the \textit{problem-space} $\mathcal{Z}$ (\ie, $z\!\in\!\mathcal{Z}$).
Then, using training samples, the ML/DL model is employed and trained as the learning-based malware detection $f:\mathcal{Z}\!\to\!\mathcal{Y}$.
That is, given an executable $z\!\in\!\mathcal{Z}$, $f$ can predict a corresponding label $y$ in the \textit{label-space} $\mathcal{Y}$ (\ie, $y\!\in\!\mathcal{Y}\!=\!\{0,\!1\}$), such that $f(z)\!=\!y$, in which $y\!=\!0$ denotes goodware while $y\!=\!1$ denotes malware.
For ease of notations like~\cite{li2021arms,ling2023survey}, we denote the learning-based malware detection that can return the malicious probability as $g: \mathcal{Z}\!\to\!\mathbb{R}$, in which $g(z)$ denotes the predicted malicious probability for $z$ and the opposite benign probability is naturally inferred as $1\!-\!g(z)$.

\subsection{Threat Model}
Following the widely used framework of modeling threats in adversarial machine learning~\cite{pierazzi2020intriguing}, we report our threat model in terms of goal, knowledge, and capability as follows.

\textbf{Adversary's Goal.}
Aiming at evading learning-based malware detection, it is hugely profitable for the adversary to misclassify malware as goodware, but not vice versa.
Therefore, the primary goal of the adversary is to generate an adversarial malware file $z_{adv}$ derived from an input malware $z \in \mathcal{Z}$ (\ie, $f(z) = 1$) with minimal efforts, such that $z_{adv} \in \mathcal{Z}$ can not only evade the target learning-based malware detection $f$ (\ie, $f(z_{adv}) = 0$) but also preserve the same semantics as $z$~\cite{ling2023survey}.

\textbf{Adversary's Knowledge and Capability.}
We start with an adversary who intends to perform the classic \textit{zero-knowledge black-box attack}~\cite{pierazzi2020intriguing,maiorca2019towards} against the learning-based malware detection system.
It indicates that the adversary has no prior information on the target system in terms of training data, extracted feature sets, employed learning algorithms with parameters, and model architectures with weights.
However, it should be clarified that the zero-knowledge black-box attack retains some minimal information about the target system, including the specific detection task (\eg, static or dynamic analysis), the employed feature type that represents executables (\eg, image, sequence, or graph), and the querying feedback.
In this paper, we restrict the adversary's knowledge to only knowing that our target system is based on static analysis that focuses on advanced abstract graph representations like {\cfg}, and knowing the predicted label $f(z)$ along \textit{with} or \textit{without} its probability $g(z)$ after inputting $z$.
Furthermore, to ensure that realistic adversarial malware files can be generated, the adversary has the capability of manipulating Windows executables while adhering to its standard specifications~\cite{pe_format_2022}.
\section{Design of \mytoolsc}\label{sec:method}

\subsection{Problem Formulation}\label{subsec:attack_formulation}
Targeting learning-based Windows malware detection, the primary goal of this paper is to generate a realistic adversarial malware file $z_{adv} \in \mathcal{Z}$ from a given malware $z$, such that $z_{adv}$ is not only misclassified as goodware but also maintains the same semantics as $z$.
To this end, we explore the problem-space attack~\cite{pierazzi2020intriguing} that applies \textit{semantics-preserving transformations} within the problem space that can transform $z$ to $z_{adv}$ step-by-step.
In particular, as formulated in \Eqref{equation:formulation_with}, we start to define the first black-box attack scenario with the malicious probability $g(\cdot)$ by reducing the probability that $z_{adv}$ is predicted as malicious as much as possible.
\begin{align}
    \argmax_{\mathbb{\mathbf{T}}} & \ g(z) - g(z_{adv}) \ \blacktriangleright \text{with predicted probabilities}\label{equation:formulation_with} \\
    \argmin_{\mathbb{\mathbf{T}}} & \ f(z_{adv}) \quad\quad \blacktriangleright \text{without predicted probabilities}\label{equation:formulation_without} \\ 
    \text{s. t.:} \quad & f(z) = 1, ~f(z_{adv}) = 0, ~z_{adv} = \mathbb{\mathbf{T}}(z) \in \mathcal{Z} \\
    & \mathbf{T} = T_{1} \circ T_{2} \circ \cdots \circ T_{n} \in \mathbb{T} \label{equation:formulation_trans}
\end{align}
in which $T \in \mathbb{T}$ denotes one of the atomic transformations that can transform one executable into another semantics-preserving executable;
$\mathbf{T}=T_{1} \circ \cdots \circ T_{n}$ denotes a finite and ordered sequence of $n$ transformations that $z$ can step-by-step transform $z$ into an adversarial malware, \ie, $z_{adv} = \mathbb{\mathbf{T}}(z)$.

Similarly, \Eqref{equation:formulation_without} defines another stricter black-box setting where the adversary can only obtain the predicted label $f(\cdot)$ without any predicted probabilities, which simply minimizes the predicted label of $z_{adv}$ to 0 since 0 denotes goodware.

\subsection{Overview Framework of \mytoolsc}\label{subsec:method:overview}
\begin{figure}[htbp]
    \centering
    \setlength{\abovecaptionskip}{3pt}
    \setlength{\belowcaptionskip}{1pt}
    \includegraphics[width=0.98\columnwidth,keepaspectratio]{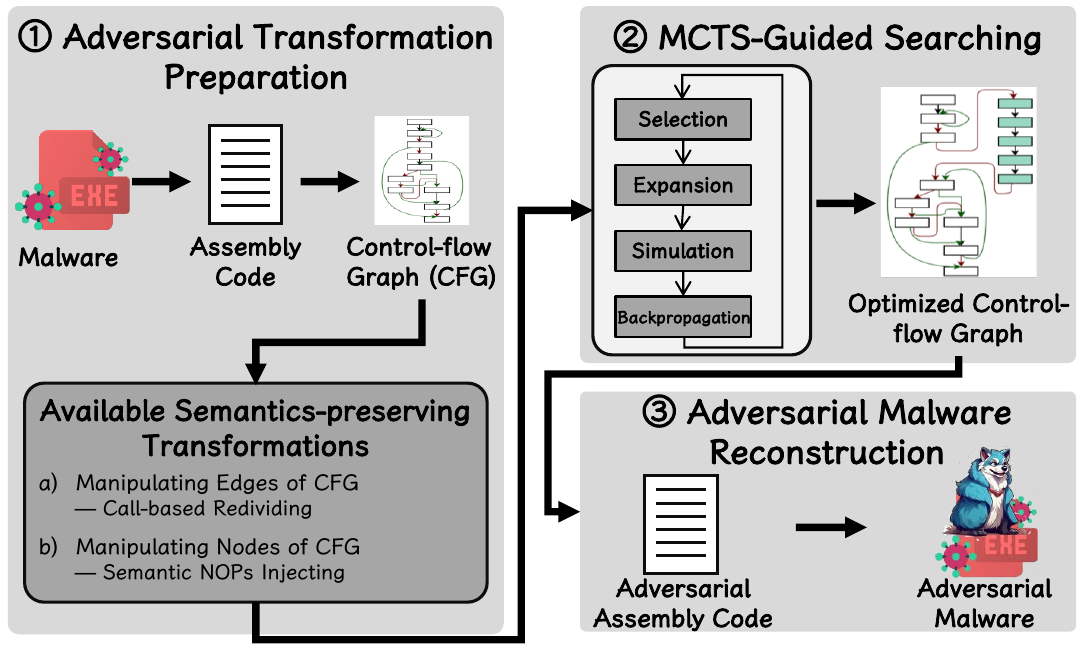}
    \caption{The overview framework of \mytool.}
    \label{figure:attack_framework}%
\end{figure}

As depicted in \Figref{figure:attack_framework}, the overview framework of \mytool mainly consists of three backbone phases, \ie, \ding{192} adversarial transformation preparation, \ding{193} MCTS-guided searching, and \ding{194} adversarial malware reconstruction, which are elaborated in the following parts of \Secref{subsec:method:transformation}, \Secref{subsec:method:searching}, and \Secref{subsec:method:reconstruction}, respectively.

\subsubsection{Adversarial Transformation Preparation}\label{subsec:method:transformation}
To prepare adversarial transformations that manipulate the given malware while preserving its semantics, we initially represent a malware sample as the {\cfg}, in which each node denotes a basic block and each edge denotes a control-flow path between two basic blocks during execution.
The main reason to represent malware with {\cfg} is twofold.
i) {\cfg} encapsulates the intrinsic control flows during execution, containing rich semantic and structural information of assembly instructions.
Therefore, if we can manipulate both {\cfg}'s nodes and edges, we can change both the semantics and structures for generating possible adversarial malware files in a more fine-grained fashion, thereby making them less noticeable by possible defenders.
ii) {\cfg} has gained widespread adoption in various software analysis tasks~\cite{ling2020multilevel}, and {\cfg}-based malware detection is extensively proven to be technically advanced and highly effective in both industry and academia~\cite{yan2019classifying,ling2022malgraph,herath2022cfgexplainer}.
Therefore, if we could successfully attack these advanced {\cfg}-based malware detection as representative cases, it would demonstrate the state-of-the-art attack effectiveness of {\mytool}.

However, if we transform one executable into another by directly manipulating its {\cfg}, it is extremely easy to cause various unexpected issues like addressing or processing errors, rendering the transformed executable unable to execute properly or even crashing immediately.
To circumvent these issues and avoid being noticeable by potential defenders, we propose a novel semantics-preserving transformation for Windows executables, namely {\call}.
It \textit{redivides} the basic blocks that contain at least one ``$\mathsf{call}$'' instruction for concurrently manipulating both nodes and edges in the {\cfg}.

To be specific, for a given executable, the {\call} first identifies and annotates all available basic blocks with ``$\mathsf{call}$'' instruction(s).
Then, supposing there is a basic block $\mathbf{v}$ containing a ``$\mathsf{call}$'' instruction, it takes the ``$\mathsf{call}$'' as the dividing line, and attempts to redivide the original basic block $v$ into a combination of three consecutive basic blocks (\ie, the fore-basic-block $\mathbf{v}_{fore}$, the post-basic-block $\mathbf{v}_{post}$, and the mid-basic-block $\mathbf{v}_{mid}$).
Finally, to avoid the basic block of $\mathbf{v}_{mid}$ being easily noticed by defenders due to having only two instructions of ``$\mathsf{call}$'' and ``$\mathsf{jmp}$'', {\call} enriches the assembly instructions in $\mathbf{v}_{mid}$ by injecting semantic {\nop}s~\cite{Christodorescu2005SemanticsawareMD,lucas2021malware} before the ``$\mathsf{call}$'' or between the ``$\mathsf{call}$'' and ``$\mathsf{jmp}$'', and employs the context-free grammar from~\cite{lucas2021malware} to generate the semantic {\nop}s diversely.

\begin{figure}[htbp]
    \centering
     \begin{subfigure}{0.16\textwidth}
         \centering
         \includegraphics[width=\textwidth]{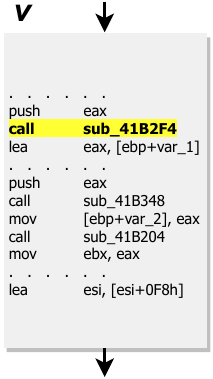}
         \caption{Before transformation.}
         \label{fig:one_basic_block:before}
     \end{subfigure}
     \hfill
     \begin{subfigure}{0.30\textwidth}
         \centering
         \includegraphics[width=\textwidth]{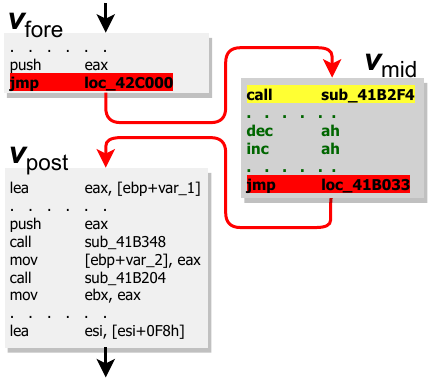}
         \caption{After applying a {\call}.}
         \label{fig:one_basic_block:after}
     \end{subfigure}
     \caption{The {\call} redivides one basic block in the ``LockBit 3.0'' ransomware (\ie, \Figref{fig:one_basic_block:before}) into a composite of three consecutive basic blocks (\ie, \Figref{fig:one_basic_block:after}).}%
     \label{fig:one_basic_block}%
\end{figure}%

As illustrated in \Figref{fig:one_basic_block}, we first show one basic block of the latest ``LockBit 3.0'', an active and famous ransom gang, as a representative example in \Figref{fig:one_basic_block:before}.
After applying the transformation, \Figref{fig:one_basic_block:after} shows the transformed composite of three consecutive basic blocks, in which the ends of two basic blocks (\ie, $\mathbf{v}_{fore}$ and $\mathbf{v}_{mid}$) are two newly added ``$\mathsf{jmp}$'' instructions, and those assembly instructions between  ``$\mathsf{call}$'' and ``$\mathsf{jmp}$'' in $\mathbf{v}_{mid}$ are newly added semantic {\nop}s.

\subsubsection{MCTS-Guided Searching}\label{subsec:method:searching}
Recalling our adversarial attack formulated in \Secref{subsec:attack_formulation} and the {\call} transformation towards the {\cfg} representation (\ie, $x=\phi(z)$) of a given malware $z$ in \Secref{subsec:method:transformation}, we decompose \mytool into first finding an optimized transformation sequence $\mathbf{T}$ that consecutively transforms the original {\cfg} $x$ into an adversarial {\cfg} of $x_{adv}$ (detailed in this part of \Secref{subsec:method:searching}), and then reconstructing the final adversarial malware file (\ie, $z_{adv}$), which will be detailed in \Secref{subsec:method:reconstruction}.

In essence, the optimal solution we are solving here is an optimized sequence of transformations $\mathbf{T} = T_{1} \circ \cdots \circ T_{N}$ of length $N$, and each $T_{k} =\{\mathbb{I}^{\mathsf{call}}_{k}, \mathbb{I}^{\snops}_{k}\}$ involves two decision-making processes:
i) Selecting one of all available ``$\mathsf{call}$'' instructions to be redivided, \ie, $\mathbb{I}^{\mathsf{call}}_{k}$, and it should be noted that $\mathbb{I}^{\mathsf{call}}_{k}$ can be \textit{repeatedly} selected in a recursive manner;
ii) Determining the proper semantic {\nop}s to be injected, namely $\mathbb{I}^{\snops}_{k}$, and it can be generated \textit{infinitely} by the employed context-free grammar~\cite{lucas2021malware}.
In short, determining an optimal $\mathbf{T}$ in {\mytool} requires exploring and optimizing in an \textit{infinite} and \textit{discrete} space with a limited computational budget under the \textit{black-box} setting.
To this end, our key idea is to employ an MCTS-guided searching algorithm~\cite{coulom2006efficient} for two major reasons.
First, MCTS has been widely and successfully used to solve the long-standing challenging problem of computer {Go}~\cite{coulom2006efficient,silver2016mastering} and other difficult optimization problems that require little or no domain knowledge~\cite{browne2012survey}.
Second, our task of optimizing $\mathbf{T}$ under the black-box setting is strictly limited to searching in an infinite and discrete space without prior knowledge.
Therefore, we argue that an MCTS-guided searching algorithm aligns well with our task requirements.

\begin{algorithm}[!tbp]
\small
\caption{MCTS-Guided Searching Algorithm.}
\label{algorithm:mcts}
\SetAlgoLined
\SetKwBlock{Begin}{Begin}{}
\SetKwInOut{Input}{Input}%
\SetKwInOut{Output}{Output}%
\SetKwFor{For}{for}{do}{}
\SetKwComment{Comment}{\text{\small{//}}}{}
\SetCommentSty{}

\Input{a given malware $z$ with its {\cfg} $x$, target system $f$, max length $N$, simulation number $S$,  budget $C$.}
\Output{the transformation sequence $\mathbf{T}$.} 
\Begin{
        $\mathbb{I}^{\mathsf{call}}$ $\leftarrow$ $\mathtt{GetAllCalls}(x)$\;
        $v,\ \mathbf{T} \leftarrow \mathtt{InitMCTSRootNode}(x, \mathbb{I}^{\mathsf{call}}),\ \emptyset$ \Comment*[r]{initialize}
        \For(\Comment*[f]{loop upto maximum length}){$i \leftarrow 1$ to $N$}{
            \For(\Comment*[f]{loop upto computation budget}){$j \leftarrow 1$ to $C$}{ 
                \uIf(\Comment*[f]{\scriptsize{avoid unlimited expansion}}){$\mathtt{random}\!(\!0\!,\!1\!)\!<\!0.5\!$}{$v_{selected}$ $\leftarrow$ $\mathtt{Selection}(v)$\;}
                \uElse{$v_{selected}$ $\leftarrow$ $\mathtt{Expansion}(v)$\;}
                $reward$ $\leftarrow$ $\mathtt{Simulation}(v_{selected},f,S)$\;
                $\mathtt{BackPropagation}(v_{selected}, reward)$\;
            }
           $v_{node} \leftarrow \mathtt{ChildWithHighestReward}(v)$\;
           $\mathbf{T} \leftarrow \mathbf{T}.\mathtt{append}(v_{node}.T)$\;
           $x_{adv} \leftarrow v_{node}.x$\;
           \uIf{$\mathtt{Evaded}(f,x_{adv})\!\!==\!\!True$}{
                \Return{$\mathbf{T}$}
           }%
           $v \leftarrow v_{node}$
        }
    }%
\end{algorithm}%

Algorithm~\ref{algorithm:mcts} presents the MCTS-guided searching algorithm, which inputs a given malware $z$ with its {\cfg} $x$ and outputs the transformation sequence $\mathbf{T}$.
Firstly, we obtain all available instructions $\mathbb{I}^{\mathsf{call}}$ from $x$ and initialize the MCTS's root node $z$ and $\mathbf{T}$(\textit{line 2--3}).
Meanwhile, we limit the maximum length of the optimized transformation sequence to $N$ (\textit{line 4--17}) and limit the maximum number of iterations of MCTS to $C$, \ie, the computational budget (\textit{line 5--11}).
As for the MCTS optimization process, we follow the four standard steps (\ie, $\mathtt{Selection}$, $\mathtt{Expansion}$, $\mathtt{Simulation}$, and $\mathtt{Backpropagation}$) (\textit{line 6--11}).
It is noted that, as {\call} can be performed unlimitedly, the game tree of MCTS can be unlimitedly expanded downwards, \ie, $\mathtt{Expansion}$.
Therefore, we force to select the most \textit{promising} child node (\ie, $\mathtt{Selection}$) in the established game tree via a simple random sampling (\textit{line 6}).
After $C$ iterations, we can thus obtain the child node $v_{node}$ with the highest reward value, append its transformation into the transformation sequence $\mathbf{T}$, and update the adversarial {\cfg} $x_{adv}$ (\textit{line 12--14}).
Finally, if $x_{adv}$ evades $f$, return the transformation sequence $\mathbf{T}$, otherwise, continue to use $v_{node}$ as the root node for the next round until reaching the maximum length $N$.
For simplicity, more implementation details can be found in Appendix~\ref{appendix:method:malguise}.

\subsubsection{Adversarial Malware Reconstruction}\label{subsec:method:reconstruction}
Finally, we reconstruct the adversarial malware $z_{adv} = \mathbf{T}(z)$ based on the original malware file $z$ and the optimized transformation sequence $\mathbf{T}$, which is briefly outlined in Algorithm~\ref{algorithm:reconstruction2}.
It is noted that each transformation $T_{k} =\{\mathbb{I}^{\mathsf{call}}_{k}, \mathbb{I}^{\snops}_{k}\}$ and we denote $A^{\mathsf{call}}_{k}$ as the address of $ \mathbb{I}^{\mathsf{call}}_{k}$.
We first calculate the space required for all {\call} transformations in $\mathbf{T}$ as $\Delta$ (\textit{line 2--6}).
Meanwhile, let $S_{slack}$ and $A_{slack}$ denote the size and address of the slack space in the ``.text'' section of $z$, respectively.
Similarly, let $A_{last}$ and $S_{last}$ denote the address and size of the last section.
If $\Delta$ is less than the size of slack space, we directly take $A_{slack}$ as the injecting address $A^{inj}$(\textit{line 7--8}).
Otherwise, we will add a new section, whose starting injection address is computed as $A^{inj}=A_{last}\!+\!\texttt{RoundUp}(S_{last}, page\_size)$(\textit{line 9--10}).
This is because, according to the standard Windows executable specifications~\cite{pe_format_2022}, its section size must be a multiple of the architecture's page size (\ie, $4KB$ for x86 and MIPS), to prevent unexpected issues (\eg, addressing errors) that may arise during execution.
Apart from this, other subsequent actions should be taken to meet the Windows specifications, such as setting the size for the ``.text'' section or the newly added section and adjusting other fields (\eg, ``size of image'') in the header of Windows executables.

\begin{algorithm}[tbp]
\small
\caption{Adversarial Malware Reconstruction}
\label{algorithm:reconstruction2}
\SetAlgoLined
\SetKwBlock{Begin}{Begin}{}
\SetKwFor{For}{for}{do}{}%
\SetKwFunction{Patch}{Patch}\SetKwFunction{GetSize}{GetSize}\SetKwFunction{AddSection}{AddSection}\SetKwFunction{RoundUp}{RoundUp}
\SetKwInOut{Input}{Input}%
\SetKwInOut{Output}{Output}%
\SetKwComment{Comment}{\texttt{//}}{}
\SetCommentSty{}
\Input{original malware $z$, the transformation sequence $\mathbf{T}$.}
\Output{the adversarial malware $z_{adv}$.} 
\Begin{
    $\Delta \leftarrow 0$ \;
    \For{$k \leftarrow 1$ \KwTo $N$} {
        $\Delta \leftarrow \Delta + \GetSize(\mathbb{I}^{\mathsf{call}}_{k}) + \GetSize(\mathbb{I}^{\snops}_{k})$\;
        $\mathbb{I}^{\mathsf{jmp}}_{k} \iff \mathsf{jmp} \text{\textvisiblespace} [A^{\mathsf{call}}_{k}+\GetSize(\mathbb{I}^{\mathsf{call}}_{k})]$\;
        $\Delta \leftarrow \Delta + \GetSize(\mathbb{I}^{\mathsf{jmp}}_{k})$ \;
    }
    \uIf(\Comment*[f]{inject into the slack space}){$\Delta < S_{slack}$}
    {$A^{inj} \leftarrow A_{slack}$}
    \uElse(\Comment*[f]{inject into a new section}){
    $A^{inj} \leftarrow A_{last} + \RoundUp(S_{last}, page\_size)$ \;
    }
    take actions to meet the standard specifications\;
    $z_{adv}\!\leftarrow\!\mathtt{Adv\_Patch}(z, \mathbf{T}, A^{inj})$ \Comment*[l]{refer to Algorithm~\ref{algorithm:adversarial:patch}}
    
    \Return{$z_{adv}$.}%
}%
\end{algorithm}

After obtaining the injecting address $A^{inj}$, we reconstruct $z_{adv}$ with the procedure of $\mathtt{Adv\_Patch}$, which is mainly outlined Algorithm~\ref{algorithm:adversarial:patch}.
In particular, we first replace the selected $\mathbb{I}^{call}_{k}$ instruction with a new $\mathsf{jmp}$ instruction (\ie, ``$\mathsf{jmp} \text{\textvisiblespace} [A^{inj}_{k}]$''), which transfers the control-flow to the injecting address for the $k$-th transformation in the new destination, \ie, $A^{inj}_{k}$ (\textit{line 4}).
Secondly, starting from the injecting address $A^{inj}_{k}$ of either the slack space or the newly added section, we deposit $\mathbb{I}^{call}_{k}$ and $\mathbb{I}^{\snops}_{k}$ in order (\textit{line 5--7}).
Subsequently, we inject a new $\mathsf{jmp}$ instruction (\ie, $\mathbb{I}^{\mathsf{jmp}}_k\!\iff\!\mathsf{jmp} \text{\textvisiblespace} [A^{\mathsf{call}}_{k}+\mathsf{GetSize}(\mathbb{I}^{\mathsf{call}}_{k})]$), which transfers the control-flow backward to the next instruction of $\mathbb{I}^{call}_{k}$ in the original section  (\textit{line 8--9}).
Finally, after all of $N$ transformations are applied, we can reconstruct the final adversarial malware $z_{adv}$ that preserves the same semantics as the original malware $z$.

\begin{algorithm}[tbp]
\small
\caption{Procedure of $\mathtt{Adv\_Patch} (z, \mathbf{T}, A^{inj})$}
\label{algorithm:adversarial:patch}
\SetAlgoLined
\SetKwBlock{Begin}{Begin}{}
\SetKwFor{For}{for}{do}{}%
\SetKwComment{Comment}{\texttt{//}}{}
\SetCommentSty{}
\SetKwFunction{Patch}{Patch}\SetKwFunction{GetSize}{GetSize}\SetKwFunction{AddSection}{AddSection}
\SetKwFunction{RoundUp}{RoundUp}
\SetKwInOut{Input}{Input}%
\SetKwInOut{Output}{Output}%
\Begin{
    $z_{adv}, \ A^{inj}_1 \leftarrow z, \ A^{inj}$\;
    \For{$k \leftarrow 1$ \KwTo $N$}{
        
        $z_{adv} \leftarrow \Patch(z_{adv},\ A^{call}_k,\ \mathsf{jmp} \text{\textvisiblespace} [A^{inj}_{k}])$ \;

        $z_{adv} \leftarrow \Patch(z_{adv},\ A^{inj}_k,\ \mathbb{I}^{call}_k)$ \;
        $A^{inj}_k \leftarrow A^{inj}_k + \GetSize(\mathbb{I}^{call}_k)$\;

        $z_{adv} \leftarrow \Patch(z_{adv},\ A^{inj}_k,\ \mathbb{I}^{\snops}_k)$ \;
        $A^{inj}_k \leftarrow A^{inj}_k + \GetSize(\mathbb{I}^{\snops}_k)$\;

        $z_{adv} \leftarrow \Patch(z_{adv},\ A^{inj}_k,\ \mathbb{I}^{\mathsf{jmp}}_k)$ \;
        
        $A^{inj}_{k+1} \leftarrow A^{inj}_k + \GetSize(\mathbb{I}^{\mathsf{jmp}}_k)$\;
    }
    \Return{$z_{adv}$}%
}%
\end{algorithm}

\begin{figure}[htb]
    \centering
    \setlength{\abovecaptionskip}{2pt}
    \setlength{\belowcaptionskip}{2pt}
    \includegraphics[width=0.90\columnwidth,keepaspectratio]{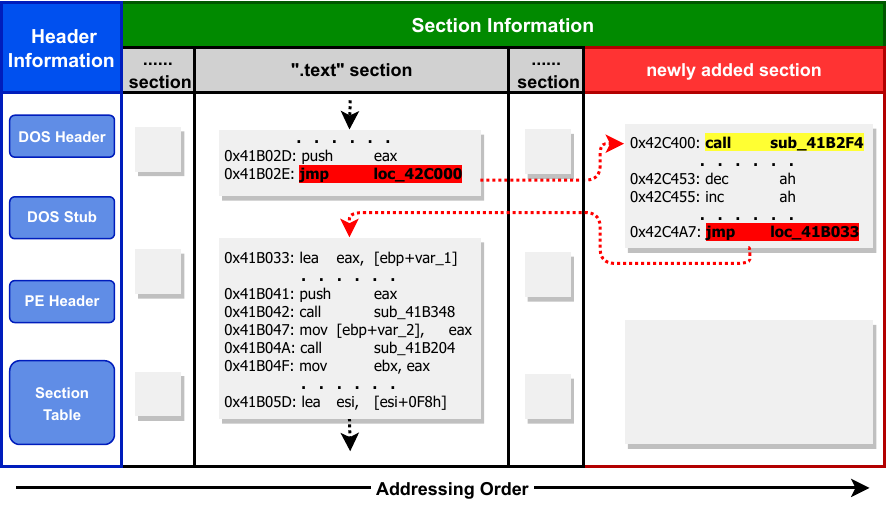}
    \caption{The conceptual layout of the reconstructed adversarial Windows malware file for the ``LockBit 3.0'' ransomware.}
    \label{figure:malware_reconstruction}
\end{figure}

\Figref{figure:malware_reconstruction} shows the conceptual layout of the generated adversarial ``LockBit'' after taking one {\call} transformation.
It is important to note that, just applying two optimized transformations of {\call} to ``LockBit 3.0'', the generated adversarial ``LockBit'' can successfully evade three state-of-the-art learning-based malware detection (\ie, MalGraph, Magic, and MalConv) and the famous anti-virus product of Kaspersky in our evaluation.
\section{Evaluation}\label{sec:evaluation}
This section is dedicated to conducting evaluations aiming at answering the following five research questions:
\begin{itemize}[leftmargin=10pt,labelwidth=3pt,labelsep=2pt,itemindent=0pt,itemsep=0pt,topsep=0pt,parsep=0pt]
    \item \textbf{RQ1 (Attack Performance)}: What is the attack performance of {\mytool} against the state-of-the-art learning-based Windows malware detection systems?

    \item \textbf{RQ2 (Impacting Factors)}: What impacting factors affect the attack performance of {\mytool}?
    
    \item \textbf{RQ3 (Utility Performance)}: Does the adversarial malware generated by {\mytool} maintain the original semantics?
    
    \item \textbf{RQ4 (Real-world Performance)}: To what extend does {\mytool} evade existing commercial anti-virus products?

    \item \textbf{RQ5 (Possible Defenses)}: What is the attack performance of {\mytool} against potential defenses?

\end{itemize}

\subsection{Evaluation Setup}

\subsubsection{Benchmark Dataset}\label{evaluation:dataset}
We utilize the same benchmark dataset as employed in prior studies~\cite{ling2022malgraph}.
This dataset is a balanced wild dataset of 210,251 Windows executables with 108,610 goodware and 101,641 malware, consisting of 848 different malware families.
As summarized in Table~\ref{tab:dataset_statistics}, we split it into three disjoint training/validation/testing sets and more details are in  Appendix~\ref{appendix:dataset}.

\begin{table}[htbp]
    \centering
    \small
    \caption{Summary statistics of the benchmark dataset.}
    \label{tab:dataset_statistics}
    \begin{tabular}{ccccc}
    \toprule
    \textbf{Dataset}  & Training    &   Validation  &  Testing   & Total \\
    \midrule
    Malware           & 81,641      &   10,000      &   10,000      & 101,641  \\
    Goodware          & 88,610      &   10,000      &   10,000      & 108,610  \\
    Total             & 170,251     &   20,000      &   20,000      & 210,251  \\
    \bottomrule
    \end{tabular}%
\end{table}%

\subsubsection{Target Systems with Detecting Performance}\label{evaluation:target_systems}
We evaluate the attack performance of {\mytool} against two kinds of target systems (\ie, learning-based malware detection systems and real-world anti-virus products) as follows.

$\bullet$ \textbf{Learning-based Windows malware detection systems.}
We first employ three SOTA learning-based malware detection (\ie, MalGraph~\cite{ling2022malgraph}, Magic~\cite{yan2019classifying}, MalConv~\cite{raff2017malware}) from either top-tier academic conferences or highly cited publications.
To avoid possible biases, we directly adopt their publicly available implementations with default hyper-parameters and evaluate their detecting performance with three commonly used metrics, \ie, AUC, TPR/FPR, and balanced Accuracy (bACC).
Table~\ref{tab:targetmalwaredetection} shows that all of them show similar detecting performance as presented in their original publications, affirming their excellent performance in detecting malware.

\begin{table}[htb]
  \centering
  \small
  \renewcommand\tabcolsep{3.3pt}
  \caption{The detecting performance of three learning-based Windows malware detection systems in our testing dataset.}
    \begin{tabular}{cccccc}
    \toprule
    \multirow{2}[2]{*}{\tabincell{c}{Target\\Systems}} & \multirow{2}[2]{*}{\tabincell{c}{AUC\\(\%)}} & \multicolumn{2}{c}{FPR = 1\%} & \multicolumn{2}{c}{FPR = 0.1\%} \\
    \cmidrule{3-6}          &       & TPR (\%)   & bACC (\%)  & TPR (\%)   & bACC (\%) \\
    \midrule
    MalGraph       &   99.94    &     99.34  &     99.18   &   92.78   &   96.36   \\
    Magic          &   99.89    &     99.02  &     99.02   &   89.28   &   94.59   \\
    MalConv        &   99.91    &     99.22  &     99.12   &   86.54   &   93.22   \\
    \bottomrule
    \end{tabular}%
  \label{tab:targetmalwaredetection}%
\end{table}

$\bullet$ \textbf{Anti-virus products.}
We additionally employ five anti-virus products, \ie, McAfee, Comodo, Kaspersky, ClamAV, and Microsoft Defender ATP (MS-ATP)~\cite{Microsoft_Defender_ATP_website}, due to their widespread recognition in the security community of Windows malware detection.
In particular, McAfee, Comodo, and Kaspersky are three award-winning commercial anti-viruses recommended in~\cite{pcmag_website}.
ClamAV~\cite{clamav_website} is the most popular open-sourced anti-virus engine, which has been extensively employed in both academia and industry.
MS-ATP~\cite{Microsoft_Defender_ATP_website} is a learning-based security protection tool for Windows and is awarded a perfect 5-star rating by~\cite{ms_atp_award}.
Notably, we do not employ VirusTotal as it strongly advises against using their anti-viruses for comparative evaluations~\cite{virustotal_not_use_2}.

\subsubsection{Baseline Attacks}\label{evaluation:baseline_attacks}
We compare {\mytool} with two kinds of baseline attacks, \ie, adversarial attacks and obfuscations.
In principle, to facilitate fair comparisons, we follow the same evaluation settings of all baseline attacks as their publications or implementations.

$\bullet$ \textbf{Adversarial attacks}:
We employ two SOTA adversarial attacks:
i) MMO~\cite{lucas2021malware} is a white-box adversarial attack that uses gradient-based optimizations to guide binary diversification tools to manipulate the raw bytes of Windows malware.
Its maximum iteration number is 200 and the increment rate of adversarial malware size is limited to 5\%.
ii) SRL~\cite{zhang2022semantics} is a black-box adversarial attack that employs reinforcement learning to iteratively inject semantic {\nop}s into the malware's {\cfg}.
Its maximum iteration number, the injection budget, and the modified basic blocks are set to be 30, 5\%, and 1250, respectively.
Notably, {SRL} only generates the adversarial {\cfg} features rather than realistic malware files, making it incompatible with MalConv, which requires raw bytes as inputs.

$\bullet$ \textbf{Obfuscations}:
We employ three obfuscation tools that have been widely employed to obfuscate Windows executables and follow their default settings for evaluations.
In particular, UPX~\cite{upx2020} is an open-source and universal packing tool for executables by performing compression on the entire file; 
VMProtect~\cite{vmprotect_document} typically uses code virtualization for obfuscations via simulating a virtual machine executing the key part in executables;
Enigma~\cite{enigma} employs a combination of multiple obfuscation techniques, \eg, import table elimination, API simulation, code virtualization, \etc.

\subsubsection{Evaluation Metrics}\label{evaluation:metrics}
We employ two kinds of evaluation metrics as follows.

$\bullet$ \textbf{Attack Success Rate (ASR)}.
ASR is the most common evaluation metric for adversarial attacks~\cite{ling2019deepsec,zhang2022semantics,lucas2021malware}.
Given a candidate malware set $\mathbf{Z}$, ASR is defined as the ratio of generated adversarial malware that successfully evades the target system (\ie, $f(z_{adv})\!=\!0$) among all malware (\ie, $f(z)\!=\!1$).
\begin{equation}
ASR = \frac{|(f(z)=1) \land (f(z_{adv})=0)|}{|f(z)=1|}, \  \forall z \in \mathbf{Z}
\label{equation:attack:success:rate}
\end{equation}
where $|\cdot|$ counts the number that meets the condition.

$\bullet$ \textbf{Semantics Preservation Rate (SPR)}.
As the generated adversarial malware might not preserve the same semantics as the original malware, \ie, it cannot be executed or lose the original malicious behaviors.
To this end, we define SPR as the ratio of adversarial malware with the original semantics preserved among all adversarial malware as follows.
\begin{equation}
SPR = \frac{|Sem(z, z_{adv}) = 1|}{|(f(z)=1) \land (f(z_{adv})=0)|}, \  \forall z \in \mathbf{Z}
\label{equation:semantic:preservation:rate}
\end{equation}
where $Sem(z, z_{adv})\!=\!1$ denotes $z_{adv}$ and $z$ maintain the same semantics.
\Secref{subsubsec:malicious:preserve} will detail how to measure SPR empirically.

\subsubsection{Implementation Details}\label{evaluation:implementation}
{\mytool} is primarily implemented with Python and evaluated on a computer equipped with 20 Intel Xeon CPUs, 128 GB memory, and 4 NVIDIA GeForce RTX 3090.
Firstly, {\mytool} uses IDAPython in IDA Pro 6.4~\cite{ida_pro} to disassemble Windows executables and represent them as {\cfg}s.
In the MCTS-guided searching algorithm, by default, we set the max length $N$ to 6, set the computational budget $C$ to 40, set the simulation number $S$ to 1, and limit the size of injected semantic {\nop}s to no more than 5\% of the original size.
To reconstruct the adversarial malware file, we mainly employ two Python libraries, pefile\footnote{pefile:\url{https://github.com/erocarrera/pefile}} and LIEF\footnote{LIEF:\url{https://lief-project.github.io}}, to parse and patch Windows executables.
It is worth noting that, those injected semantic {\nop}s can be generally divided into four categories: arithmetic (\eg, ``$\mathsf{add\text{\textvisiblespace}eax,\,1;\ sub\text{\textvisiblespace}eax,\,1}$''), logical (\eg, ``$\mathsf{add\text{\textvisiblespace}eax,\,eax}$''), comparison (\eg, ``$\mathsf{cmp\text{\textvisiblespace}eax,\,eax}$''), and data transfer (\eg, ``$\mathsf{push\text{\textvisiblespace}eax;\ pop\text{\textvisiblespace}eax}$'').
\subsection{Evaluation Results \& Analysis}

\subsubsection{\textbf{Answer to RQ1 (Attack Performance)}}
\label{subsubsec:attack:performance:learning}
To assess the attack performance of {\mytool}, we evaluate it by comparing its ASR performance with all baseline attacks on all 10,000 testing malware samples from the benchmark dataset, which is illustrated in Table~\ref{tab:ASRofMalGuiseAndBaseline_add_with_obfuscations}.
Recall in~\Secref{evaluation:baseline_attacks}, MMO is a white-box adversarial attack, serving as an upper bound to assess the attack performance of its black-box attacks.
All three obfuscations only apply to the \textit{w/o prob.} scenario as they do not require any feedback from the target system.

\begin{table}[htbp]
\setlength\tabcolsep{1.4pt}
  \centering
  \small
  \caption{The ASR performance (\%) comparisons between \mytool and baseline attacks against three target systems under two black-box scenarios, \ie, \textit{w/} \textit{prob.} and \textit{w/o} \textit{prob.}}
    \begin{tabular}{ccrrrrrr}
    \toprule
    \multirow{2}[6]{*}{\tabincell{c}{Black-box\\Scenarios}} & \multirow{2}[6]{*}{Attacks} & \multicolumn{2}{c}{MalGraph} & \multicolumn{2}{c}{Magic} & \multicolumn{2}{c}{MalConv} \\
    \cmidrule(r{2pt}){3-4} \cmidrule(l{2pt}){5-6} \cmidrule(l{2pt}){7-8} & & \tabincell{c}{FPR\\=1\%} & \tabincell{c}{FPR\\=0.1\%} & \tabincell{c}{FPR\\=1\%} & \tabincell{c}{FPR\\=0.1\%} & \tabincell{c}{FPR\\=1\%} & \tabincell{c}{FPR\\=0.1\%} \\
    \midrule
    \multirow{3}[4]{*}{\tabincell{c}{\textit{w/}\\\textit{prob.}}}    & MMO     & 15.55   & 52.30 & 12.82 & 40.13 & 11.99 & 39.66 \\
    & SRL   & 2.39    & 19.59 & 25.38 & 86.77 & —      & —       \\
    \cmidrule{2-8}                  & \mytool                         & \textbf{97.47}   & \textbf{97.77} & \textbf{99.29} & \textbf{99.42} & \tabincell{c}{\textbf{34.36}\\(\textbf{97.76})} & \tabincell{c}{\textbf{97.38}\\(\textbf{99.77})}\\
    \midrule
    \multirow{6}[6]{*}{\tabincell{c}{\textit{w/o}\\\textit{prob.}}}   & MMO  & 3.73     & 27.83     & 3.41     & 25.46     & 2.46     & 20.72 \\
    & SRL  & 2.59     & 15.28     & 3.84     & 47.48     & —     & — \\
    \cmidrule{2-8}
    & UPX       & 0.55  & 4.43  & 3.30  & {39.80}     & 0.31  & 9.32 \\
    & VMProtect & 0     & 0     & 0.23  & 4.33      & 0     & 0    \\
    & Enigma    & 0.81  & 11.69 & 0     & 28.96     & 0     & 0.24 \\
    \cmidrule{2-8}
    & \mytool & \textbf{96.84}     & \textbf{96.49}     & \textbf{99.27}     & \textbf{99.07}     & \tabincell{c}{\textbf{31.41}\\(\textbf{95.18})}   & \tabincell{c}{\textbf{88.02}\\(\textbf{99.77})} \\
    \bottomrule
    \end{tabular}%
  \label{tab:ASRofMalGuiseAndBaseline_add_with_obfuscations}
  \scriptsize{``—'' means SRL does not apply to MalConv as it cannot generate real malware files.}
\end{table}

For all attacks, Table~\ref{tab:ASRofMalGuiseAndBaseline_add_with_obfuscations} shows that lower FPR values for the three target systems correspond to higher ASRs achieved by each attack.
The reason is evident that a lower value of FPR allowed by the binary classifier indicates it has a higher threshold.
Hence, the adversary can more easily reduce the predicted malicious probability to a level below the threshold.
Additionally, it is evident that each attack performs no better in the scenario of \textit{w/o prob.} than in the scenario of \textit{w/ prob.}

\textbf{Adversarial attacks}:
{MMO} achieves low ASR performance against all three target systems in both attack scenarios.
Particularly, in the case of FPR=1\% for all target systems, the ASRs of MMO are below 16\% and 4\% in the scenarios of \textit{w/ prob.} and \textit{w/o prob.}, respectively.
These imply that, although MMO theoretically can be used against all target systems, it shows inferior attack performance against them.
The main reason we conjecture is that, since MMO manipulates the entire raw bytes of malware with binary diversification techniques in general, it does not take into account the discriminative features employed in different target systems.

For {SRL}, in both scenarios of \textit{w/ prob.}  and \textit{w/o prob.}, it shows obviously higher ASRs against Magic than MalGraph in both FPRs.
This is mainly because SRL is specifically designed to attack Magic which purely builds on {\cfg}.
However, the hierarchical nature of MalGraph that combines both the function call graph and {\cfg} further weakens the attack performance of SRL, as SRL only manipulates nodes of {\cfg} and neglects to manipulate its edges.

\textbf{Obfuscations}:
It is observed that all three obfuscations show inferior attack performance, which once again validates that traditional obfuscation tools are largely ineffective against learning-based malware detection~\cite{raff2017malware,kim2022obfuscated,aghakhani2020malware}.
Specifically, VMProtect achieves the worst attack performance against all three target systems, while UPX and Enigma show slightly better attack performance, but remain unsatisfactory with all achieved ASRs not exceeding 40\%.
This is mainly because, VMProtect typically obfuscates only a small portion of the malware file~\cite{xu2018vmhunt,vmprotect_document}, while both UPX and Enigma can globally obfuscate the entire malware file, which slightly increases the likelihood of altering the discriminative semantic features.
However, as UPX and Enigma remain unknown without the security experts' intervention, it is still challenging to purposefully manipulate the discriminative semantic features.
More evaluations in terms of size alteration ratio are in Appendix~\ref{appendix:section:more:evaluations}.

\textbf{Compare {\mytool} with all baseline attacks}:
Table~\ref{tab:ASRofMalGuiseAndBaseline_add_with_obfuscations} shows that {\mytool} achieves the best attack performance against all three target systems in all scenarios and cases.
When attacking MalGraph and Magic in all scenarios and cases, all ASR values achieved by \mytool exceed 97\%.
More importantly, even in the strict attack scenario of \textit{w/o prob.}, \mytool still maintains its ASR performance nearly unchanged, \ie, decrease by no more than 1\%, compared to the scenario of \textit{w/ prob.}
When attacking MalConv, although the ASR performance of \mytool is much better than that of all baseline attacks, its ASR is still relatively poor (\ie, below 35\%) in the case of FPR=1\%.
However, when investigating {\mytool} in the subsequent \Secref{subsubsec:ablation:study}, we find that its attack performance against MalConv is highly dependent on the injected semantic {\nop}s.
Thus, by incorporating the 25 most frequently used semantic {\nop}s, \mytool is improved to achieve higher ASRs of 97.76\% and 95.18\% against MalConv in both \textit{w/ prob.} and \textit{w/o prob.} scenarios, respectively.

\begin{tcolorbox}[size=title,boxsep=1pt,left=1pt,right=1pt,top=0pt,bottom=0pt]{
\textbf{Answer to RQ1 (Attack Performance)}:
Prior obfuscations and adversarial attacks either cannot provide satisfactory attack effectiveness or fail to scale well to different types of learning-based Windows malware detection.
However, even in the strict black-box attack scenario (\textit{w/o prob.}), {\mytool} is agnostic to learning-based Windows malware detection with a high attack success rate exceeding 95\% in most cases.
}%
\end{tcolorbox}

\subsubsection{\textbf{Answer to RQ2 (Impacting Factors)}}
\label{subsubsec:ablation:study}
We conduct a series of ablation studies to explore the impacting factors that affect the attack performance of \mytool.

\textbf{Impact of key parts in {\call}}.
The core of {\mytool} is to apply the transformation of {\call}, which primarily involves two key parts: $\dag$ \underline{injecting semantic {\nop}s} and $\ddag$ \underline{redividing $\mathsf{call}$ instructions}, for manipulating the {\cfg}'s \underline{nodes} and \underline{edges}, respectively.
To examine the impact of the above two parts, we compare the ASR performance of \mytool with its two variants that apply only one of the key parts (\ie, \mytoolNOSpace$^{\dag}$ and \mytoolNOSpace$^{\ddag}$) under the attack scenario of \textit{w/ prob.}

From Table~\ref{tab:ablation:twokeycomponent}, both {\mytoolNOSpace$^{\dag}$} and {\mytoolNOSpace$^{\ddag}$} show almost negligible attack performance (with ASRs below 5\%) against both MalGraph and Magic in both FPRs.
When attacking MalConv, {\mytoolNOSpace$^{\ddag}$} still shows inferior attack performance, while {\mytoolNOSpace$^{\dag}$} shows better attack performance, achieving ASRs of 23.04\% and 79.86\% in the cases of FPR=1\% and FPR=0.1\%, respectively.
The reason is evident that the essence of {\mytoolNOSpace$^{\dag}$} is to inject semantic {\nop}s into malware, which can significantly and directly alter their raw bytes.
Thus, {\mytoolNOSpace$^{\dag}$} can easily affect the detecting performance of MalConv which inputs the raw bytes of malware.
Finally, comparing with the above two variants, {\mytool} achieves considerably higher attack performance with over 97\% ASR in all cases.
More evaluations against anti-viruses are in Appendix~\ref{appendix:section:more:evaluations}.
All the above observations indicate that \mytool with only manipulating either {\cfg}'s nodes or edges demonstrates inferior attack performance, and only by combining both can the attack performance of \mytool be maximized.
\begin{table}[htbp]
  \centering
  \setlength\tabcolsep{0.5pt}
  \small
  \caption{The ASR performance (\%) of three \mytool variants against three target systems under the scenario of \textit{w/ prob.}}
    \begin{tabular}{lcccccc}
    \toprule
    \multicolumn{1}{c}{\multirow{2}[4]{*}{\tabincell{c}{\mytool\\Variants}}} & \multicolumn{2}{c}{MalGraph} & \multicolumn{2}{c}{Magic} & \multicolumn{2}{c}{MalConv}  \\
    \cmidrule(r{1pt}){2-3} \cmidrule(l{1pt}){4-5} \cmidrule(l{1pt}){6-7}  \multicolumn{1}{c}{} & \tabincell{c}{\footnotesize{FPR=1\%}} & \tabincell{c}{\footnotesize{FPR=0.1\%}} & \multicolumn{1}{c}{\tabincell{c}{\footnotesize{FPR=1\%}}} & \multicolumn{1}{c}{\tabincell{c}{\footnotesize{FPR=0.1\%}}} & \multicolumn{1}{c}{\tabincell{c}{\footnotesize{FPR=1\%}}} & \multicolumn{1}{c}{\tabincell{c}{\footnotesize{FPR=0.1\%}}}\\
    \midrule
    {\mytoolNOSpace$^{\dag}$}  & 0.02  & 4.35 & 1.70  &  3.34 & 23.04 & 79.86 \\
    {\mytoolNOSpace$^{\ddag}$}  & 0.10  & 4.87 & 0.10  &  1.42 & 0.17  &  1.86 \\
    {\mytoolNOSpace} & \bf 97.47  & \bf 97.77 & \bf 99.29 & \bf 99.42 & \bf \tabincell{c}{34.36\\97.76} & \bf \tabincell{c}{97.38\\99.77} \\
    \bottomrule
    \end{tabular}
  \label{tab:ablation:twokeycomponent}
\end{table}

\begin{figure}[htbp]
    \centering
    \begin{subfigure}{0.15\textwidth}
     \centering
     \includegraphics[width=\textwidth]{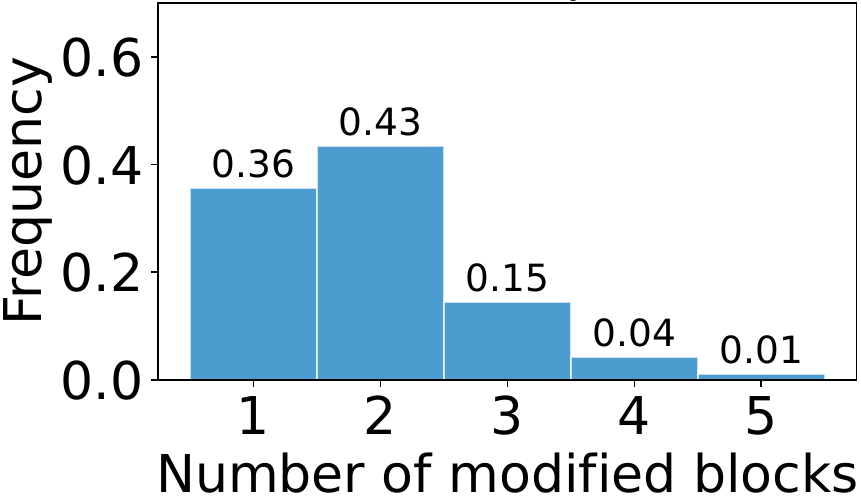}
     \caption{MalGraph}
     \label{subfig:impact:malgraph}
    \end{subfigure}
    \hspace{-1.8mm}
    \begin{subfigure}{0.15\textwidth}
     \centering
     \includegraphics[width=\textwidth]{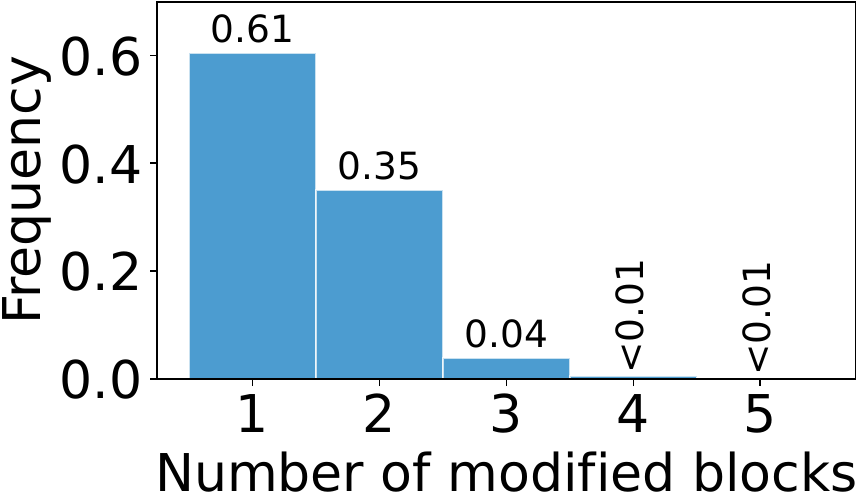}
     \caption{Magic}
     \label{subfig:impact:magic}
    \end{subfigure}
    \hspace{-1.8mm}
    \begin{subfigure}{0.15\textwidth}
     \centering
     \includegraphics[width=\textwidth]{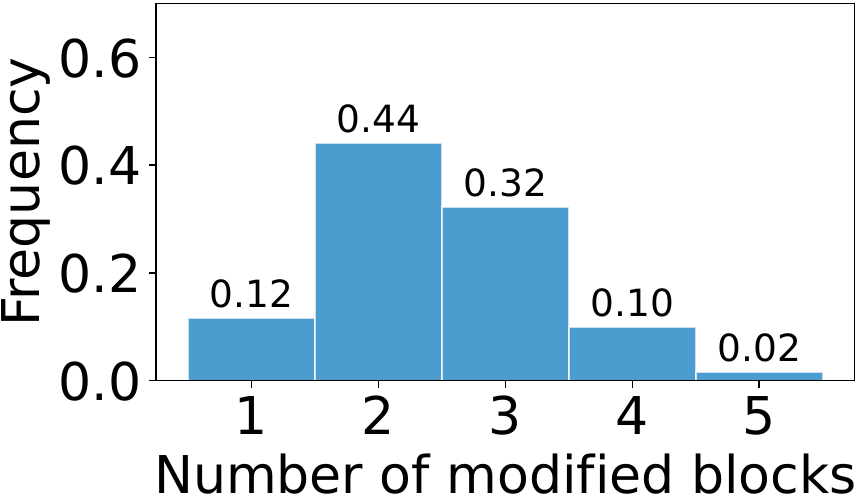}
     \caption{MalConv}
     \label{subfig:impact:malconv}
    \end{subfigure}
    \caption{Frequency of the number of modified basic blocks of all adversarial malware that evades the three target systems.}
    \label{fig:impact:modified:blocks}
\end{figure}

\textbf{Impact of the number of modified basic blocks in {\cfg}}.
We investigate the impact of the number of modified basic blocks for all generated adversarial malware that successfully evade the three learning-based malware detection when FPR=1\%, and show their corresponding frequencies in \Figref{fig:impact:modified:blocks}.
It is observed that over 98\% of all adversarial malware files that successfully evade the target systems only require modifying no more than four basic blocks in its {\cfg}.
In particular, \mytool only needs to modify one basic block to make 61\% of malware samples evade Magic, and only modify two basic blocks to make 79\% and 56\% of malware samples evade MalGraph and MalConv, respectively.
These evaluations indicate that, by only modifying a small number of basic blocks in {\cfg}, {\mytool} could offer excellent attack performance against all three learning-based malware detection systems.

\begin{figure}[htb]
    \centering
    \includegraphics[width=0.65\columnwidth,keepaspectratio]{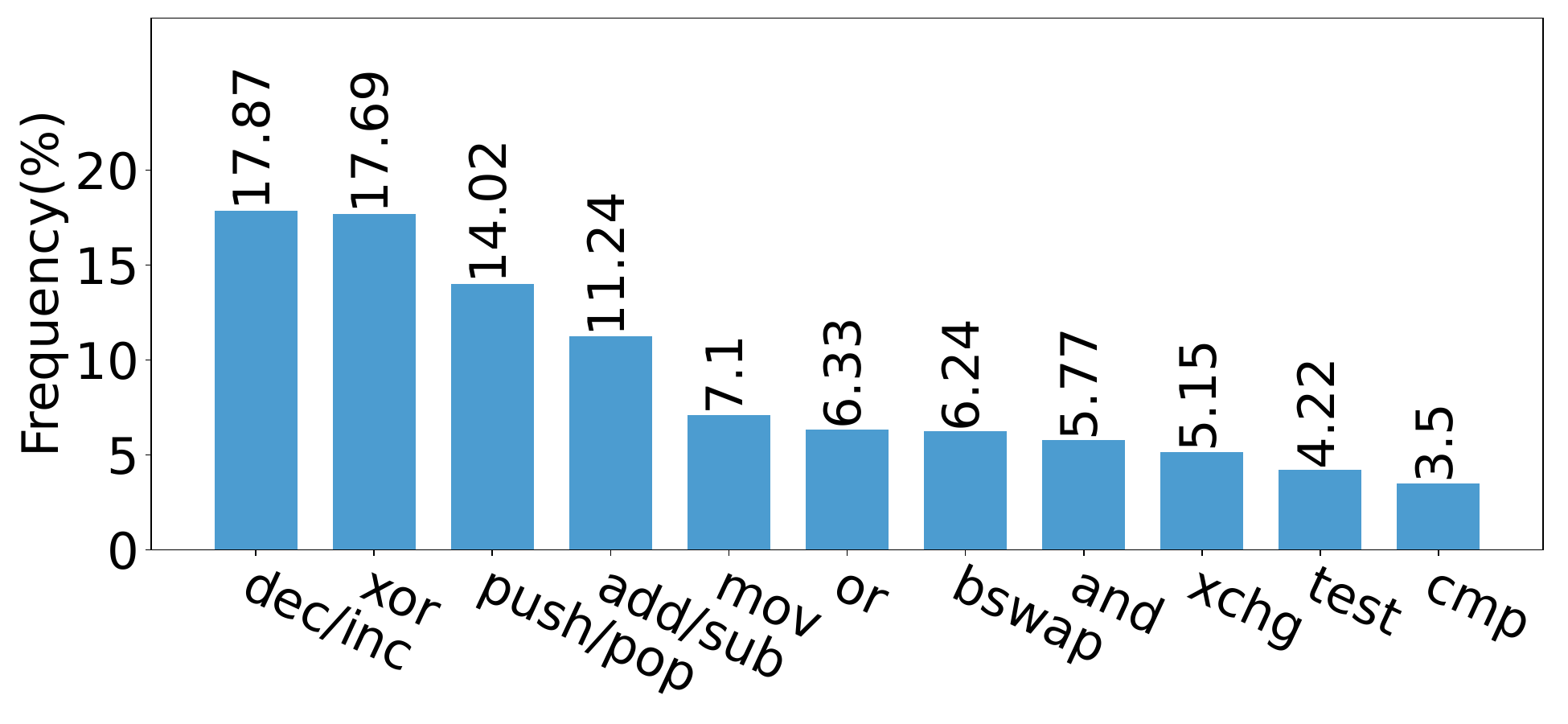}
    \caption{Occurrence frequency of different opcodes in semantic {\nop}s that lead to successful evasions against MalConv.}
    \label{figure:occurrence_frequency_of_opcodes}
\end{figure}

\textbf{Impact of different types of semantic {\nop}s}.
When performing {\mytool} against MalConv, we examine the impact of different opcodes in the injected semantic {\nop}s by presenting the occurrence frequency of different opcodes that lead to successful evasions in \Figref{figure:occurrence_frequency_of_opcodes}.
It is observed that, some opcodes (\eg, \texttt{dec/inc}, \texttt{xor}) occur with a relatively high frequency, while some others (\eg, \texttt{cmp} and \texttt{test}) occur with a lower frequency.
Next, we limit {\mytool} to use the 25 most frequently used opcodes, it is observed from Table~\ref{tab:ASRofMalGuiseAndBaseline_add_with_obfuscations} that the ASR performance of {\mytool} ASR against MalConv has significantly increased, highlighting the importance of different types of semantic {\nop}s in {\mytool}.
To sum up, we can conclude that the attack performance of {\mytool} can be significantly improved by fine-tuning the types of semantic {\nop}s to be injected.
More evaluations about the impact of the size of semantic {\nop}s for {\mytool} are discussed in Appendix~\ref{appendix:section:more:evaluations}.

\begin{figure}[htb]
    \centering
    \begin{minipage}{0.22\textwidth}
    \centering
    \includegraphics[width=\columnwidth,keepaspectratio]{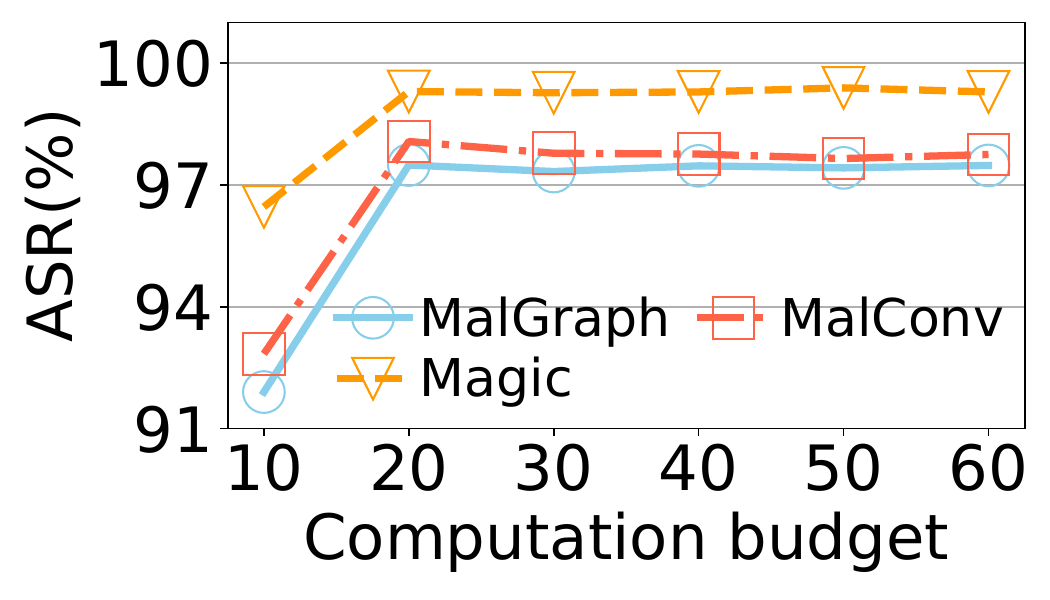}
    \caption{Impact of the computation budget $C$.}
    \label{figure:impact:computation:budget}
    \end{minipage}%
    \hfill
    \begin{minipage}{0.22\textwidth}
    \centering
    \includegraphics[width=\columnwidth,keepaspectratio]{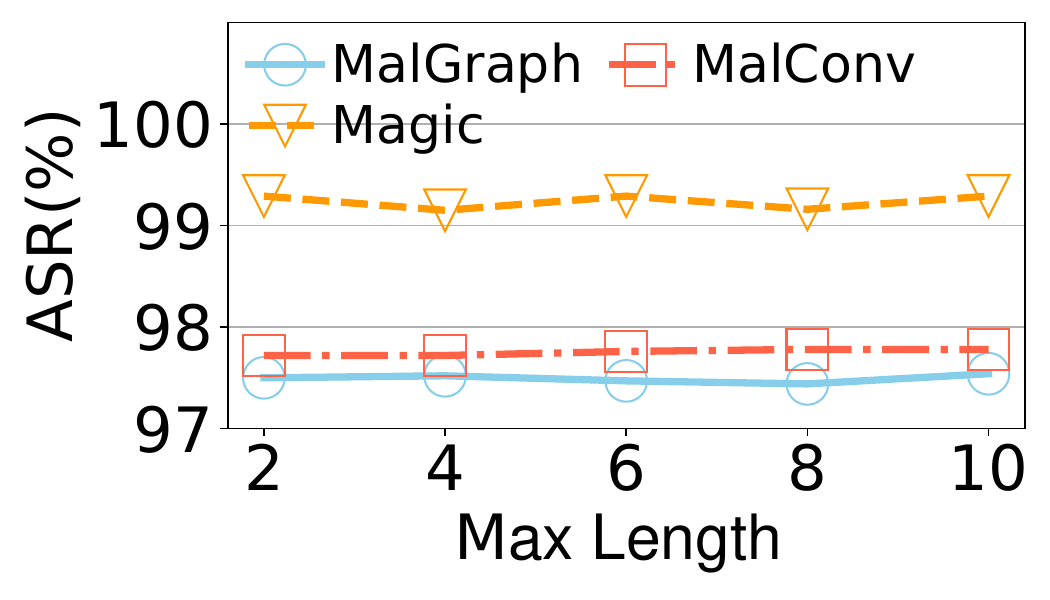}
    \caption{Impact of the max length $N$.}
    \label{figure:impact:mcts:level}
    \end{minipage}%
\end{figure}

\textbf{Impact of the hyper-parameters in MCTS.}
We investigate the impact of the hyper-parameters in MCTS (\ie, $C$ and $N$ in Algorithm~\ref{algorithm:mcts}) for {\mytool}.
First, for the computation budget $C$, we follow the same implementation settings detailed in \Secref{evaluation:implementation} and only vary $C$ to 10, 20, 30, 40, 50, and 60.
From \Figref{figure:impact:computation:budget}, it is observed that, as $C$ increases from 10 to 20, the ASRs of \mytool rise sharply up to over 97\% against all three target systems.
When $C$ reaches 20, its ASR performance tends to stabilize at a high value of over 97\%.

Similarly, for the max length $N$, we only vary $N$ (\ie, 2, 4, 6, 8, and 10) to show its impact on \mytool in \Figref{figure:impact:mcts:level}.
It is observed that when $N\!=\!2$, the overall attack performance of \mytool against all three target systems is pretty well, with ASRs exceeding 97\%.
With the increase of N, the ASRs of \mytool remain stable at a high value of over 97\% against all target systems.
These observations indicate that {\mytool} can achieve high attack performance even with no more than two transformations of {\call} employed.

\begin{tcolorbox}[size=title,boxsep=1pt,left=1pt,right=1pt,top=0pt,bottom=0pt]{
\textbf{Answer to RQ2 (Impacting Factors)}:
To sum up, targeting different learning-based malware detection systems, different hyper-parameters in {\mytool} have different impacts on the attack performance, and thereby we can fine-tune its hyper-parameters to enhance its attack performance.
}
\end{tcolorbox}

\subsubsection{\textbf{Answer to RQ3 (Utility Performance)}}\label{subsubsec:malicious:preserve}
As previously defined in \Eqref{equation:semantic:preservation:rate}, we evaluate the utility performance of {\mytool} and two baseline adversarial attacks in terms of semantics preservation rate (SPR), whose core is to judge whether the adversarial malware $z_{adv}$ has the same semantics as the original malware $z$, \ie, $Sem(z, z_{adv})$.
Due to the inherent complexity of executables, there is no exact solution to judge $Sem(z, z_{adv})$~\cite{apel2009measuring} and we resort to an empirical verification to judge it by collecting and comparing the two API sequences (\ie, $\mathtt{API}_{z_{adv}}$ and $\mathtt{API}_{z}$) invoked by both $z_{adv}$ and $z$ when they are run on the Cuckoo sandbox~\cite{cuckoo_website}.
As shown in \Eqref{equation:normalize:distance}, to quantify the semantic difference between $z_{adv}$ and $z$, we thus compute a normalized edit distance $dist_{norm} (z, z_{adv})$ between the two API sequences as follows.
\begin{equation}
    dist_{norm} (z, z_{adv})\!=\!\frac{Distance(\mathtt{API}_{z}, \mathtt{API}_{z_{adv}})}{max(l(\mathtt{API}_{z}), l(\mathtt{API}_{z_{adv}}))}\!\in\![0, 1]
    \label{equation:normalize:distance}
\end{equation}
in which $Distance(\mathtt{API}_{z}, \mathtt{API}_{z_{adv}})$ denotes the edit distance between two sequences and $l(\cdot)$ denotes the sequence length.

However, since malware may perform random actions during execution~\cite{kasama2012malware}, the API sequences collected by running the same malware $z$ \textit{twice} in the same sandbox may be different, which means $dist_{norm} (z, z)$ almost can not take the value of 0.
Therefore, we calculate the value of $Sem(z, z) \in \{0, 1\}$ by comparing the  $dist_{norm} (z, z)$ with a general distance threshold $dist_{\Delta}$.
To determine $dist_{\Delta}$, we first analyze all original malware samples in the same sandbox twice and then select the value at the 99.5-th percentile among all the corresponding $dist_{norm} (z, z)$ as $dist_{\Delta}$.
After that, as shown in \Eqref{equation:semantic:calculation}, we can finally determine whether $z_{adv}$ and $z$ have the same semantics by comparing their normalized edit distance $dist_{norm} (z, z_{adv})$ with $dist_{\Delta}$, and further evaluate $SPR$ according to \Eqref{equation:semantic:preservation:rate}.
\begin{equation}%
Sem(z, z_{adv})\!=\!
    \begin{cases}
    1 &\!\text{if} \, dist_{norm} (z, z_{adv})\!<\!dist_{\Delta} \\
    0 &\!\text{otherwise.}
    \end{cases}
    \label{equation:semantic:calculation}
\end{equation}

Due to the extreme resource and time consumption of the above evaluation process, we randomly select 10\% from all adversarial malware that successfully evades the corresponding target system in \Secref{subsubsec:attack:performance:learning} for the subsequent evaluations and present the evaluation results in Table~\ref{table:maliciousness_preservation_rates}.
Apparently, evaluating the utility performance of {SRL} is not applicable as it only generates adversarial features rather than realistic adversarial malware files.
As for {MMO}, it is observed that the achieved SPR values against three target systems are at a low level, \ie, approximately ranging from 40\% to 50\%.
It indicates, only less than 50\% of adversarial malware generated by {MMO} preserve their original semantics.
However, {\mytool} achieves the best utility performance of over 91\% SPR for all three target systems, which demonstrates the best effectiveness of {\mytool} in preserving their original semantics.

\begin{table}[htbp]
\centering
\small
\setlength\tabcolsep{0.8pt}
\caption{The SPR (\%) of \mytool and two baseline adversarial attacks against three target systems.}
\begin{tabular}{ccccccc}
    \toprule
    \multirow{2}[3]{*}{Attacks} & \multicolumn{2}{c}{MalGraph} & \multicolumn{2}{c}{Maigc} & \multicolumn{2}{c}{MalConv} \\
    \cmidrule(r{1pt}){2-3} \cmidrule(l{1pt}){4-5} \cmidrule(l{1pt}){6-7}
     & \tabincell{c}{\footnotesize{FPR=1\%}} & \tabincell{c}{\footnotesize{FPR=0.1\%}} & \tabincell{c}{\footnotesize{FPR=1\%}} & \tabincell{c}{\footnotesize{FPR=0.1\%}} & \tabincell{c}{\footnotesize{FPR=1\%}} & \tabincell{c}{\footnotesize{FPR=0.1\%}} \\
    \midrule
    MMO     & 41.8  & 49.4  & 39.6  & 39.8  & 39.2  & 50.8 \\
    SRL     & —     & —     & —     & —     & —     & — \\
    {\mytool} & \textbf{91.84} & \textbf{91.99} & \textbf{93.45} & \textbf{92.28} & \textbf{92.67} & \textbf{91.68} \\
    \bottomrule
\end{tabular}
\label{table:maliciousness_preservation_rates}
\end{table}

\textbf{Causes for 9\% failures.}
To investigate the causes behind the approximately 9\% of failures in preserving the same semantics in the above evaluations, we conduct manual inspections using IDA Pro~\cite{ida_pro} and OllyDbg~\cite{ollydbg} and reveal two primary reasons for these failures.
First, there are a few malware samples that contain \textit{overlay}, which is not part of the official Windows executable format but is usually used to perform malicious behaviors in malware.
Despite adhering to the standard Windows executable format specifications, the adversarial malware reconstruction phase in {\mytool} might affect the overlay of those few malware samples, thereby failing to be executed identically or properly.
Second, it is a standard process for Windows executables to push/pop the return address onto/off the stack when handling the ``$\mathsf{call}$'' instruction.
However, there are few malware samples that contain \textit{junk code}, which might not follow the above process, \eg, not popping the return address.
These exceptions might render {\mytool} ineffective as {\mytool} follows the standard Windows specifications to reconstruct the adversarial malware.

\begin{tcolorbox}[size=title,boxsep=1pt,left=1pt,right=1pt,top=0pt,bottom=0pt]{
\textbf{Answer to RQ3 (Utility Performance)}:
Prior adversarial attacks either only generate non-executable adversarial ``features'', or generate a large portion of adversarial malware losing their original semantics.
However, {\mytool} exhibits the best utility performance with over 91\% of generated adversarial malware files preserving their original semantics.} 
\end{tcolorbox}

\subsubsection{\textbf{Answer to RQ4 (Real-world Performance)}}\label{subsubsec:attack:performance:antivirus}
As discussed in \Secref{evaluation:target_systems}, to further understand the real security threats of \mytool against anti-virus products, we empirically evaluate \mytool against five commercial anti-virus products by measuring their ASR performance on 1,000 testing malware samples, which are randomly selected from the testing dataset.
It is noted that, the main reason for randomly selecting 1,000 testing malware samples is that all five anti-virus products are deployed remotely on another machine, and their processing and scanning speeds are much slower than the employed learning-based Windows malware detection.

\begin{table}[htbp]
  \centering
  \small
  \renewcommand\tabcolsep{2pt}
  \caption{The ASR (\%) of \mytool against five anti-viruses.}
    \begin{tabular}{ccccccc}
    \toprule
     Attacks       &   McAfee  &   Comodo  &   Kaspersky   &   ClamAV      &   MS-ATP            \\
    \midrule
    {\mytool}      &   48.81 &   36.00 &   11.29    &   31.94    &   \textbf{70.63}  \\
    {\mytoolNOSpace}(S)  &   52.49 &   36.36 &   13.36    &   32.33   &   \textbf{74.97}  \\
    Increased ASR  &   +3.68 &   +0.36 &   +2.07    &   +0.39    &   +4.34          \\
    \bottomrule
    \end{tabular}
  \label{tab:ASRofAntiVirus}
\end{table}

We observe in Table~\ref{tab:ASRofAntiVirus} that, for four (\ie, McAfee, Comodo, ClamAV, and MS-ATP) of the five evaluated anti-viruses, {\mytool} achieves ASRs of more than 30\%.
Especially for MS-ATP, the achieved ASR of {\mytool} reaches upto 70.63\%.
As discussed in \Secref{subsubsec:ablation:study}, we further limit {\mytool} to use the 25 most frequently used opcodes and denote this variant as {\mytoolNOSpace}(S) for brevity.
Compared with the default {\mytool}, {\mytoolNOSpace}(S) clearly leads to an overall improvement in the attack performance for all five anti-virus products.
In particular, {\mytoolNOSpace}(S) increases its ASR against MS-ATP by 4.34\%, implying it can be further improved by carefully fine-tuning.

\begin{figure}[htbp]
    \centering
    \includegraphics[width=0.75\columnwidth,keepaspectratio]{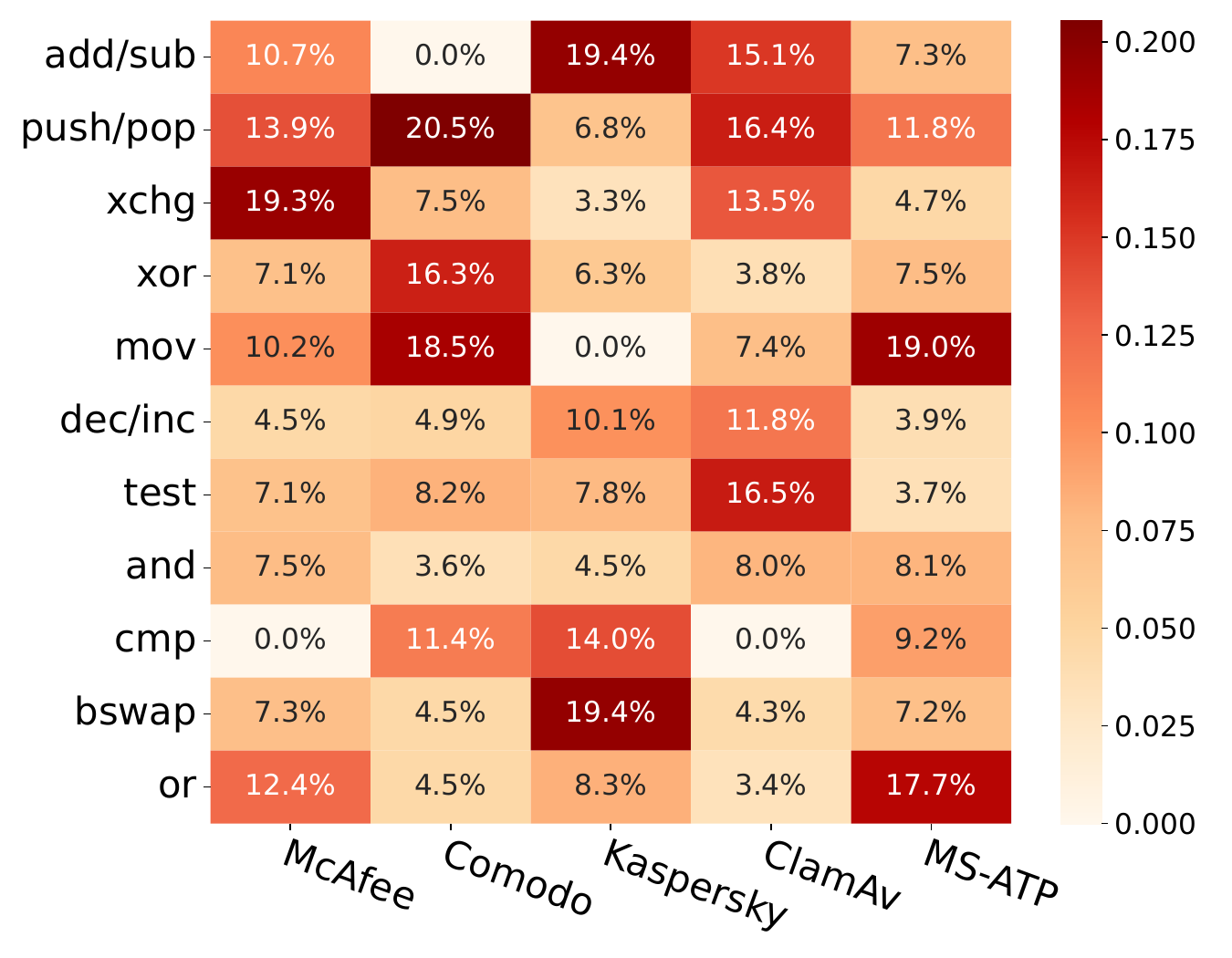}
    \caption{Occurrence frequency of different opcodes in semantic {\nop}s that lead to successful evasions for five anti-viruses.}
    \label{figure:occurrence_frequency_AVs}
\end{figure}

\textbf{Impact of different types of semantic {\nop}s.}
We further investigate how the different opcodes in the injected semantic {\nop}s affect the attack performance of {\mytool} against anti-virus products.
Similarly, by limiting {\mytool} to use the 25 most frequently used opcodes, \Figref{figure:occurrence_frequency_AVs} illustrates the occurrence frequency of different opcodes of all generated adversarial malware files that lead to successful evasions for five anti-viruses.
It is clearly observed that different anti-virus products are sensitive to different semantic {\nop}s injected by {\mytool}.

\begin{table}[htbp]
  \centering
  \small
  \renewcommand\tabcolsep{2pt}
  \caption{Distribution frequency (\%) of the number of modified blocks for adversarial malware that evades anti-virus products.}
    \begin{tabular}{crrrrr}
    \toprule
     \textbf{\# of blocks} &   McAfee  &   Comodo  &   Kaspersky   &   ClamAV      &   MS-ATP            \\
    \midrule
    1   &  96.66    & 94.28 & 88.17 &  97.54    &  38.21 \\
    2   &  4.58     &  4.71 &  9.68 &  2.05     &  42.88 \\ 
    3   &  0.76     &  1.01 &  2.15 &  0.41     &  17.35 \\ 
    4   &  0        &  0    &  0    &   0       &   1.17\\ 
    5   &  0        &  0    &  0    &   0       &   0.39\\ 
    \bottomrule
    \end{tabular}
  \label{tab:block_number_AVs}
\end{table}

\textbf{Impact of the number of modified basic blocks}.
The distribution of the number of modified blocks in the adversarial malware that successfully evades the target anti-viruses is illustrated in Table~\ref{tab:block_number_AVs}.
For all five anti-viruses, the number of modified blocks in all the adversarial malware is less than 6.
Especially for McAfee, Comodo, and ClamAV, more than 90\% of adversarial malware only need to modify one block, while the other two anti-virus products (\ie, Kaspersky and MS-ATP) only need to modify two basic blocks.
Therefore, we can conclude that \mytool can be applied against anti-virus products by only modifying very few blocks in {\cfg}.

\begin{tcolorbox}[size=title,boxsep=1pt,left=1pt,right=1pt,top=0pt,bottom=0pt]{
\textbf{Answer to RQ4 (Real-world Performance)}:
{\mytool} is empirically evaluated to be effective against five anti-virus products in the wild.
In particular, {\mytool} achieves attack success rates of over 30\% against four of them, presenting potential tangible security concerns to real-world users.
} 
\end{tcolorbox}

\subsubsection{\textbf{Answer to RQ5 (Possible Defenses)}}\label{subsubsec:possible:defense}
To understand the risks that potential artifacts in \mytool might be noticed and identified by defenders, we evaluate {\mytool} with four categories of possible defenses as follows.

(1) \textbf{Adversarial training}:
Adversarial training is recognized as one of the most effective defenses against adversarial attacks~\cite{Bai2021RecentAI}.
Therefore, we evaluate the attack performance of \mytool against the three target learning-based malware detection systems that have been defended using adversarial training (\ie, the defended systems.)
To set up, we re-train them from scratch with nearly unchanged parameter settings, except for integrating corresponding adversarial malware generated by \mytool during the training process.

\begin{figure}[htbp]
    \centering
    \small
    \begin{subfigure}{0.16\textwidth}
        \centering
        \includegraphics[width=\textwidth]{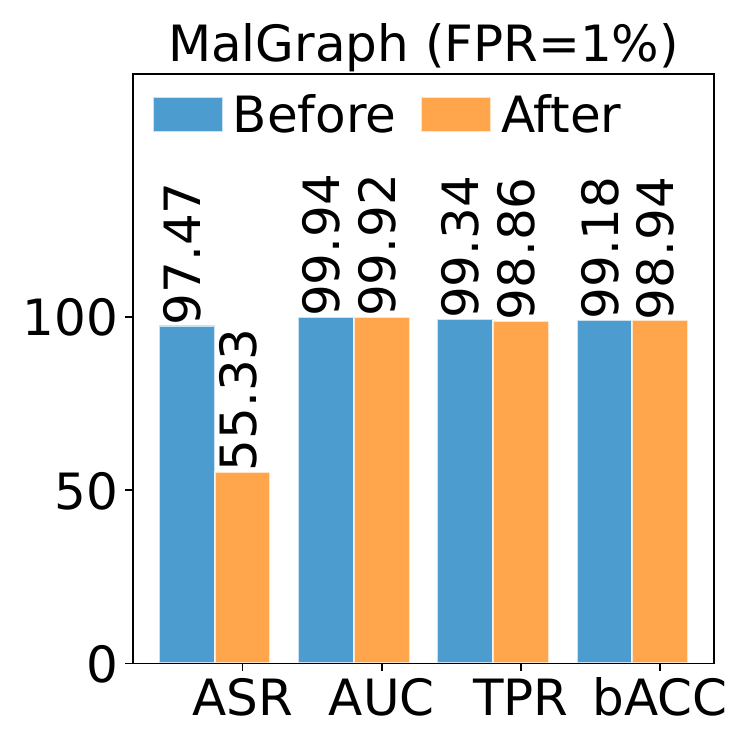}
        \caption{MalGraph}
        \label{subfig:adversarial:training:malgraph}
    \end{subfigure}
    \hspace{-2mm}
    \begin{subfigure}{0.16\textwidth}
        \centering
        \includegraphics[width=\textwidth]{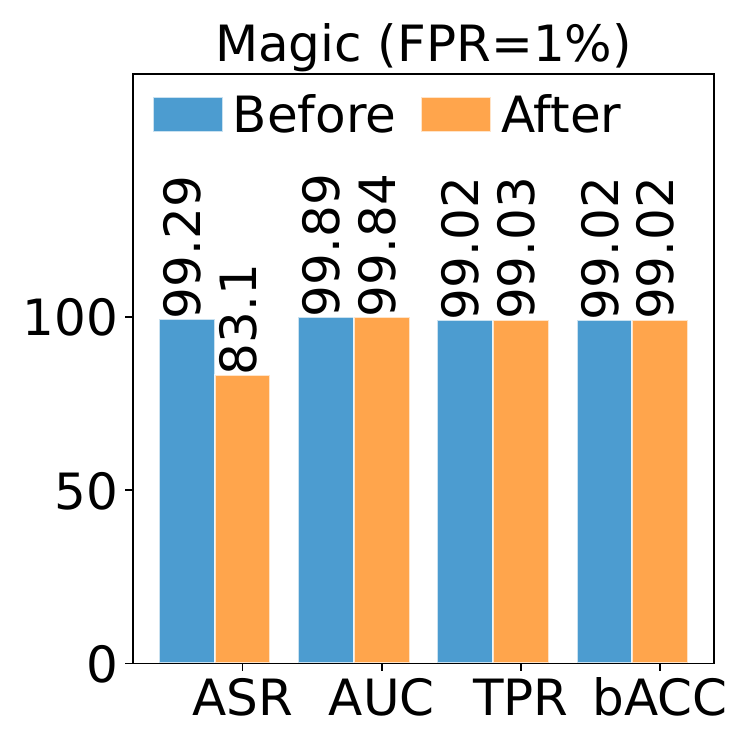}
        \caption{Magic}
        \label{subfig:adversarial:training:magic}
    \end{subfigure}
    \hspace{-2mm}
    \begin{subfigure}{0.16\textwidth}
        \centering
        \includegraphics[width=\textwidth]{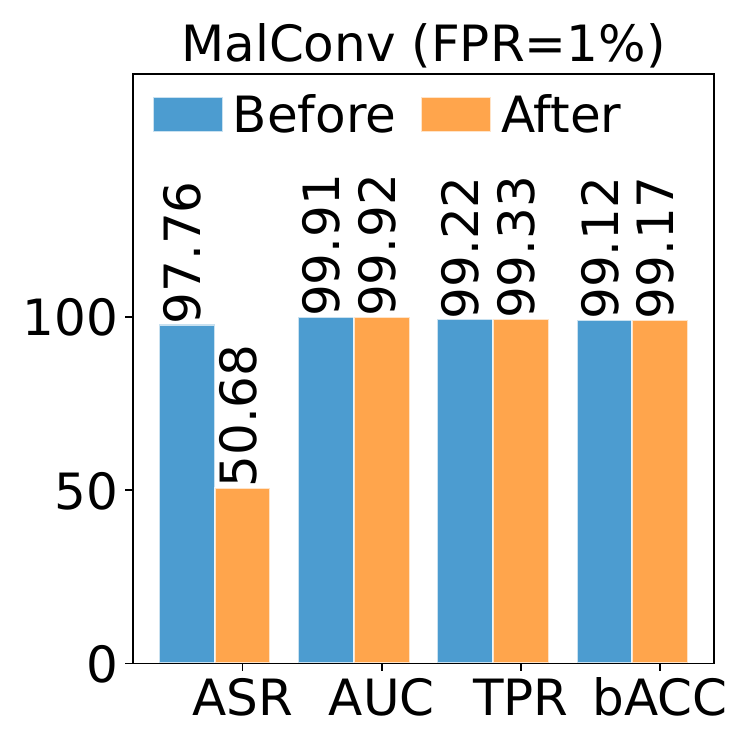}
        \caption{MalConv}
        \label{subfig:adversarial:training:malconv}
    \end{subfigure}
    \caption{The overall performance comparisons between the original (before) and the defended (after) target systems.}
    \label{fig:adversarial:training:three}
\end{figure}

\Figref{fig:adversarial:training:three} shows, after being defended by adversarial training, the detecting performance (\ie, AUC, TPR, and bACC) of the three target systems remains basically unchanged, while the corresponding ASR values achieved by \mytool have been decreased to some extent, indicating the defended systems' robustness performance (\ie, the opposite of ASR) has indeed improved.
Nonetheless, it is clearly observed that, after being defended by adversarial training, \mytool remains highly effective, achieving ASRs of 55.33\%, 83.10\%, and 50.68\% against MalGraph, Magic, and MalConv, respectively.

(2) \textbf{Binary code optimizations}:
Recalling that our proposed {\call} transformation involves injecting semantic {\nop}s and redividing ``$\mathsf{call}$'' instructions to manipulate {\cfg}, which might be noticed by potential defenders.
Therefore, to evaluate {\mytool}'s attack performance against possible defenses based on binary code optimizations (\eg, dead code removal or {\cfg} reduction), we employ the off-the-shelf IDA Pro plugin -- Optimice\footnote{\url{https://code.google.com/archive/p/optimice/}}, which has won the Hex-Rays’ IDA Pro plugin contest~\cite{OptimiceIDAProContest}.
To set up, we take Optimice's binary code optimizations as the preliminary step for the three target learning-based malware detection systems and evaluate them under the black-box scenario of \textit{w. prob.}

\begin{table}[htbp]
  \centering
  \small
  \renewcommand\tabcolsep{2.1pt}
  \caption{The ASR performance (\%) comparisons between the original target systems and the defended systems by Optimice.}
    \begin{tabular}{crccc}
    \toprule
    \multicolumn{1}{c}{\multirow{2}[4]{*}{\tabincell{c}{Target\\Systems}}} & \multicolumn{1}{c}{\multirow{2}[4]{*}{\tabincell{c}{Original\\System}}} & \multicolumn{3}{c}{the defended system by Optimice} \\
    \cmidrule{3-5}  &  & \multicolumn{1}{c}{\tabincell{c}{only \textit{w/} dead\\code removal}} & \multicolumn{1}{c}{\tabincell{c}{only \textit{w/} {\cfg}\\reduction}} & \multicolumn{1}{c}{\textit{w/} both} \\
    \midrule
    MalGraph    & 97.47   & 82.25     & 82.43   & 80.26 (-17.21) \\
    Magic       & 99.29   & 97.94     & 96.91   & 93.85 (-5.44) \\
    MalConv     & 97.76   & 80.07     & 75.15   & 74.63 (-23.13) \\
    \bottomrule
    \end{tabular}%
  \label{tab:defense:optimice}%
\end{table}%

Table~\ref{tab:defense:optimice} shows that, with either dead code removal or {\cfg} reduction, or a combination of the above two, the ASR performance achieved by \mytool has decreased, suggesting the robustness performance of the defended systems has indeed increased accordingly.
However, even when defended with binary code optimizations, \mytool still achieves relatively high attack performance, with ASRs exceeding 74\%.

(3) \textbf{Heuristic-based adversarial detection}:
To evaluate the impact of possible defenders who are fully aware of the implementation details of {\mytool}, we implement an adaptive defense of heuristic-based adversarial detection, which consists of two heuristic rules: $\mathbb{R}_1$ and $\mathbb{R}_2$.
In particular, $\mathbb{R}_1$ indicates, other than the ``.text'' section, there exists another or more sections with executable permission and fixed addresses inside.
$\mathbb{R}_2$ indicates, there exists at least one basic block that contains a ``$\mathsf{call}$'' instruction followed by a ``$\mathsf{jmp}$'' instruction with semantic {\nop}s in between.
It is noted that, as \mytool iteratively generates semantic {\nop}s with the context-free grammar~\cite{lucas2021malware}, determining whether a specific code fragment is semantic {\nop}s is well-known as non-trivial and undecidable~\cite{Chris2005ssss,Singh2018challenges} and we thus employ Optimice~\cite{OptimiceIDAProContest} to detect the semantic {\nop}s as above.
To set up, we first randomly select 15000 adversarial malware from all those that have successfully evaded three learning-based Windows malware detection systems and 15000 goodware from the benchmark dataset, and then employ the above adversarial detection to check whether the adversarial malware can be detected, and finally show the detection results in Table~\ref{tab:defense:heuristic:rules}.

\begin{table}[htbp]
  \centering
  \small
  \caption{The detecting performance (\ie, TPR and FPR) of adversarial detection based on two heuristic rules}
    \begin{tabular}{cccc}
    \toprule
    \tabincell{c}{Adversarial Detection} & TPR & FNR=(100\%-TPR) & FPR\\
    \midrule
    $\mathbb{R}_1$-based                     & 70.15 & 29.85 & 5.66  \\
    $\mathbb{R}_2$-based                     & 96.91 & 3.09  & 52.16 \\
    $\mathbb{R}_1\,\&\,\mathbb{R}_2$-based   & 67.75 & 33.25 & 2.95  \\
    \bottomrule
    \end{tabular}%
  \label{tab:defense:heuristic:rules}%
\end{table}%

It is observed that both TPR and FPR of $\mathbb{R}_1$-based adversarial detection are low, suggesting that most goodware indeed has only a ``.text'' section.
However, there is a lot of adversarial malware with only a ``.text'' section as long as the semantic {\nop}s could be injected into the remaining slack space.
Moreover, $\mathbb{R}_2$-based adversarial detection shows the best TPR but the worst FPR.
It indicates, although $\mathbb{R}_2$ can be used to detect almost all adversarial malware, there exists over 50\% of goodware is misclassified as adversarial malware.
Similar to $\mathbb{R}_1$, the adversarial detection that combines both $\mathbb{R}_1$ and $\mathbb{R}_2$ shows a TPR of 67.7\%, indicating that about 33.2\% of adversarial is misclassified as goodware, \ie, evading the corresponding the adversarial detection.
Moreover, it also shows an FPR of 2.9\%, which is higher than the upper limit of FPR (\eg, 1\% or 0.1\%) that is tolerated by general malware detection systems.

To sum up, this adaptive defense of heuristic-based adversarial detection can detect adversarial malware generated by \mytool to a certain extent as it is equipped with all the implementation details of \mytool.
However, it also shows the dilemma that TPR and FPR cannot be balanced simultaneously, \ie, failing to offer satisfactory detecting performance.

(4) \textbf{Fuzzy hashing}:
As \mytool is optimized to use as few transformations as possible to preserve the same semantics, it also enables possible defenders to use fuzzy-hashing-based malware analysis (\ie, ssdeep~\cite{ssdeep}, TLSH \cite{tlsh}, SpamSum~\cite{pyspamsum}, \etc) for defenses, which has been extensively studied~\cite{cao2022fork,yang2023recmal}.
To implement the fuzzy-hashing-based defenses, we first submit all malware samples in the training set to initialize \textit{the malware database}, and then use the validation set to determine the thresholds satisfying FPR=1\% and FPR=0.1\% for fair comparisons.
After that, we evaluate the detecting performance of fuzzy-hashing-based defenses within the testing set, which is shown in Table~\ref{tab:defense:ssdeep}.
By comparing Table~\ref{tab:targetmalwaredetection} with Table~\ref{tab:defense:ssdeep}, we find that all fuzzy-hashing-based defenses show less stable and worse detecting performance.

\begin{table}[htbp]
  \centering
  \small
  \renewcommand\tabcolsep{2pt}
  \caption{The detecting performance (\ie, AUC, TPR, bACC) and ASR performance of fuzzy-hashing-based defenses}
    \begin{tabular}{crrrrrrr}
    \toprule
    \multirow{2}[4]{*}{\tabincell{c}{Fuzzy-hashing-\\based defenses}} & \multicolumn{1}{c}{\multirow{2}[4]{*}{AUC}} & \multicolumn{3}{c}{FPR=1\%} & \multicolumn{3}{c}{FPR=0.1\%} \\ \cmidrule(r{2pt}){3-5} \cmidrule(l{2pt}){6-8}  \multicolumn{1}{c}{} &       & \multicolumn{1}{c}{TPR} & \multicolumn{1}{c}{bACC} & \multicolumn{1}{c}{ASR} & \multicolumn{1}{c}{TPR} & \multicolumn{1}{c}{bACC} & \multicolumn{1}{c}{ASR} \\
    \midrule
    ssdeep  & 83.92 & 57.61 & 78.70 & 52.94 & 10.16 & 55.31 & 100 \\
    TLSH    & 74.23 & 55.55 & 77.66 & 73.70 & 42.27 & 71.44 & 100 \\
    SpamSum & 83.98 & 57.72 & 78.75 & 50.22 & 38.91 & 69.46 & 98.13 \\
    \bottomrule
    \end{tabular}%
  \label{tab:defense:ssdeep}%
\end{table}

We further reinforce the three fuzzy-hashing-based defenses by submitting all malware samples in the testing set to the \textit{the malware database}, indicating all these defenses have full knowledge of the original malware samples and can use fuzzy hashing for detection.
Finally, we evaluate the ASR performance of \mytool against those reinforced defenses, which is shown in the column named ``ASR'' in Table~\ref{tab:defense:ssdeep}.
It is clearly observed that, in the case of FPR=1\%, the ASRs against all three reinforced defenses exceed 50\%, and the ASRs are nearly 100\% for FPR=0.1\%, exhibiting a high attack success rate against fuzzy-hashing-based defenses.

\begin{tcolorbox}[size=title,boxsep=1pt,left=1pt,right=1pt,top=0pt,bottom=0pt]{
\textbf{Answer to RQ5 (Possible Defenses)}:
\mytool remains exceptionally effective against all three categories of possible defenses even though they are adaptively equipped with the knowledge of {\mytool}.
}
\end{tcolorbox}
\section{Discussions}\label{sec:discuss}
\subsection{Related Work}

Almost all existing studies on adversarial attacks against malware detection predominantly focus on learning-based malware detection with static features.
Owing to the vast diversity of feature representations employed by different kinds of learning-based malware detection, researchers have proposed different adversarial attacks tailored to these detection methods.
For instance, to attack those malware detection methods based on API calls, a line of adversarial attacks has been proposed via adding irrelevant API calls, which are selected by gradient-based optimizations or greedy algorithms~\cite{hu2017generating,chen2017adversarial,al2018adversarial,verwer2020robust}.
However, these adversarial attacks are impractical because they generate adversarial API calls rather than realistic executable malware files.
As for malware detection based on raw bytes, especially for MalConv~\cite{raff2017malware}, researchers have explored either partially modifying specific regions or globally modifying all raw bytes while preserving the same semantics.
For instance, all of \cite{anderson2017evading,kreuk2018deceiving,suciu2019exploring,kolosnjaji2018adversarial,demetrio2020adversarial} rely on appending or injecting maliciously generated bytes at specific locations of the input malware, while {MMO}~\cite{lucas2021malware} globally manipulates its raw bytes with binary diversification techniques.
However, we argue that those adversarial attacks are strictly limited to raw-bytes-based malware detection, and thus cannot be scalable to other malware detection.

More recently, studies have begun to explore adversarial attacks against more advanced malware detection based on abstract graph representations.
In 2022, Zhang~\etal~\cite{zhang2022semantics} proposed {SRL} against {\cfg}-based malware detection, which sequentially injects semantic {\nop}s into the {\cfg} guided by reinforcement learning.
Likewise, our work also aims to evade advanced malware detection based on {\cfg}s, but fundamentally differs from {SRL} in three key aspects.
\begin{itemize}[leftmargin=10pt,labelwidth=6pt,labelsep=4pt,itemindent=0pt,itemsep=0pt,topsep=0pt,parsep=0pt]
    \item \textbf{Fine-grained transformations}.
    {SRL} employs a coarse-grained transformation that only manipulates nodes of {\cfg}.
    However, we propose a finer-grained transformation of {\call} that manipulates both nodes and edges, making it less noticeable to potential defenders.
    
    \item \textbf{Not adversarial ``features''}.
    As discussed in \Secref{subsubsec:malicious:preserve}, {SRL} is a feature-space adversarial attack that essentially generates adversarial ``features'', while our {\mytool} can successfully generate real adversarial malware files for evasions.

    \item \textbf{Attacking generalizability}.
    Existing adversarial attacks like {SRL} are primarily evaluated with limited learning-based malware detection models, while {\mytool} additionally targets real-world anti-virus products in practice, thereby demonstrating better attacking generalizability.
\end{itemize}


\subsection{Possible Use Cases}
We discuss possible use cases of \mytool as follows.
First, \mytool, a practical black-box adversarial attack, can complement the blue team's toolkit, addressing the limitations of traditional obfuscations in attempts to evade present-day advanced learning-based malware detection.
Second, \mytool could serve as a means for the R\&D team of anti-virus vendors to expose the underlying weaknesses of learning-based malware detection under development, and thus they can proactively and purposefully improve the robustness of anti-virus products.
Third, comprehensive and impartial testing for anti-viruses is crucial for ensuring transparency and fostering trust between users and vendors.
\mytool offers a unique opportunity for third-party independent testing organizations like AV-TEST~\cite{av_test_about} to assess anti-viruses comprehensively, given the exceptional evasion capability of adversarial malware.

\subsection{Potential Ethical Concerns}\label{subsec:ethicalconcerns}
The primary objective of this study is to assess the security risks of learning-based Windows malware detection with adversarial attacks, an approved topic with established precedence in earlier studies~\cite{pierazzi2020intriguing,zhao2021structural,lucas2021malware,zhang2022semantics,he2023efficient,ling2023survey}, which is largely motivated by the concern for potential adversaries to craft Windows malware capable of evading detection, and by highlighting the necessity for more robust learning-based Windows malware detection methods.
Even though our intent is strict about assessing the security risks of learning-based Windows malware detection with {\mytool}, we recognize the potential ethical concerns associated with our study.
Therefore, to strike a balance between avoiding potential ethical concerns and assisting the security community in enhancing the robustness of learning-based Windows detection, we limit our code sharing to \textit{verified academic researchers only}, following the precedent established by previous studies~\cite{pierazzi2020intriguing,zhao2021structural,he2023efficient}.
\section{Conclusion, Limitations and Future Work}\label{sec:limit:future:work}

This paper proposes a novel semantics-preserving transformation of {\call}, capable of concurrently manipulating both nodes and edges of the {\cfg} and further presents an adversarial attack framework of {\mytool} against learning-based Windows malware detection under the black-box scenario.
Extensive evaluations demonstrate that \mytool can not only effectively evade state-of-the-art learning-based Windows malware detection systems with an attack success rate of mostly exceeding 95\%, but also can evade five commercial anti-virus products with an attack success rate of up to 74.97\%.
We believe our study raises public awareness about the security threats posed by adversarial attacks in the domain of Windows malware detection and calls for further studies to enhance the robustness of existing Windows malware detection.
However, we recognize the limitations and outline potential future work as follows.
\begin{itemize}[leftmargin=9pt,labelwidth=6pt,labelsep=4pt,itemindent=0pt,itemsep=0pt,topsep=0pt,parsep=0pt]
    \item \textbf{Verification of semantic preservation}.
    To verify if the generated adversarial malware preserves the original semantics, we have presented an automatic empirical verification in \Secref{subsubsec:malicious:preserve} and \Eqref{equation:normalize:distance} by comparing their API sequences invoked when running on the same sandbox environment.
    While practical and reasonable, this empirical verification cannot guarantee the strict semantic equivalence between two malware in all possible environments since some malware may invoke random APIs in different environments.
    
    \item \textbf{Dynamic-analysis-based malware detection}.
    {\mytool} targets learning-based Windows malware detection, which belongs to static analysis.
    Particularly, the core of \mytool is to manipulate the {\cfg} of executables via the {\call} transformation, which does not change their execution flow and thereby does not impact dynamic analysis.
    Therefore, although dynamic analysis is less ubiquitously deployed due to excessive time and resource consumption, we leave generating adversarial malware against dynamic-analysis-based malware detection as future work.
    
    \item \textbf{Format-agnostic adversarial malware}.
    Since there is no malware analysis technique that is universally applicable to all types of malware with different file formats and operating systems, existing malware analysis normally points out the targeted file format and operating system.
    While in this paper, our {\mytool} targets the Windows malware in the file format of the portable executable, we will explore how to generate format-agnostic adversarial malware against all kinds of malware detection in future work.
\end{itemize}

\section*{Acknowledgments}
We sincerely appreciate the shepherd and anonymous reviewers for their valuable comments to improve our work.
We would also like to thank Bolin Zhou, Zhiqin Rui, and Yuhao Peng for their valuable contributions during the review process.
This work was partly supported by
the National Natural Science Foundation of China (62202457),
the Major Research plan of the National Natural Science Foundation of China (92167203),
the Open Source Community Software Bill of Materials (SBOM) Platform (E3GX310201),
and YuanTu Large Research Infrastructure.

\bibliographystyle{plain}
\bibliography{references}
\appendix
\section{More Implementation Detail in {\mytoolsc}}\label{appendix:method:malguise}
Algorithm~\ref{appendix:algorithm:selectbestchild} outlines the main procedure of $\mathtt{Selection}$.
As the possible paths in the game tree of MCTS are infinite, exploring all the nodes in the game tree will substantially increase the computational overhead.
By employing the Upper Confidence Bounds algorithm~\cite{browne2012survey}, $\mathtt{Selection}$ could select the child of the MCTS node $v$ with the highest score considering the trade-off between the visit times and the reward value (\textit{line 4-6}).

\begin{algorithm}[htb]
\small
\caption{Procedure of $\mathtt{Selection} (v)$.}
\label{appendix:algorithm:selectbestchild}
\SetAlgoLined
\SetKwBlock{Begin}{Begin}{}
\SetKwInOut{Input}{Input}%
\SetKwInOut{Output}{Output}%
\SetKwFor{For}{for}{do}{}
\SetKwComment{Comment}{\texttt{//}}{}
\SetCommentSty{}
\Begin{
    $max\_score \leftarrow 0$\;
    \For{$v_{child}$ in children of $v$} {
        $exploit \leftarrow v_{child}.reward/v_{child}.visits$\;
        $explore \leftarrow \sqrt {2\ln ({v}.visits)/{v_{child}}.visits}$\;
        $score \leftarrow exploit+\lambda \times explore$\;
        \uIf{$max\_score < score$}{
            $max\_score$ $\leftarrow$ $score$\;
            $v_{selected}$ $\leftarrow$ $v_{child}$\;
        }
    }
}%
\Return{$v_{selected}$} \Comment*[l]{the most \textit{promising} node to be explored}
\end{algorithm}

Algorithm \ref{appendix:algorithm:defaultpolicy} outlines the main procedure of $\mathtt{Simulation}$ that returns the reward $r$.
Based on the $v_{selected}$, this procedure iteratively expands the MCTS game tree until the simulation number $S$ is reached (\textit{line 3--6}).
In each iteration, the corresponding CFG representation $x$ of the expanded node is input to the target malware detection system $f$, and the returned reward is calculated as $reward=1-f(x)$ (\textit{line 6}).

\begin{algorithm}[htb]
\small
\caption{Procedure of $\mathtt{Simulation}\!(v_{selected}\!,f\!,\!S)\!$}
\label{appendix:algorithm:defaultpolicy}
\SetAlgoLined
\SetKwBlock{Begin}{Begin}{}
\SetKwInOut{Input}{Input}%
\SetKwInOut{Output}{Output}%
\SetKwFor{For}{for}{do}{}%
\SetKwBlock{Begin}{Procedure $\mathtt{Simulation}(v_{selected}, f, S)$}{}
\Begin{
    $v' \leftarrow v_{selected}$\;
    \For{$i \leftarrow 1$ \KwTo $S$} {
        $v' \leftarrow \mathtt{Expansion}(v')$\;
        $x \leftarrow v'.x$\;
    }
    $reward \leftarrow 1-f(x)$\;
}%
\Return{$reward$}%
\end{algorithm}

\section{More about the Benchmark Dataset}\label{appendix:dataset}
\begin{figure}[htb]
    \centering
    \includegraphics[width=0.45\textwidth,keepaspectratio]{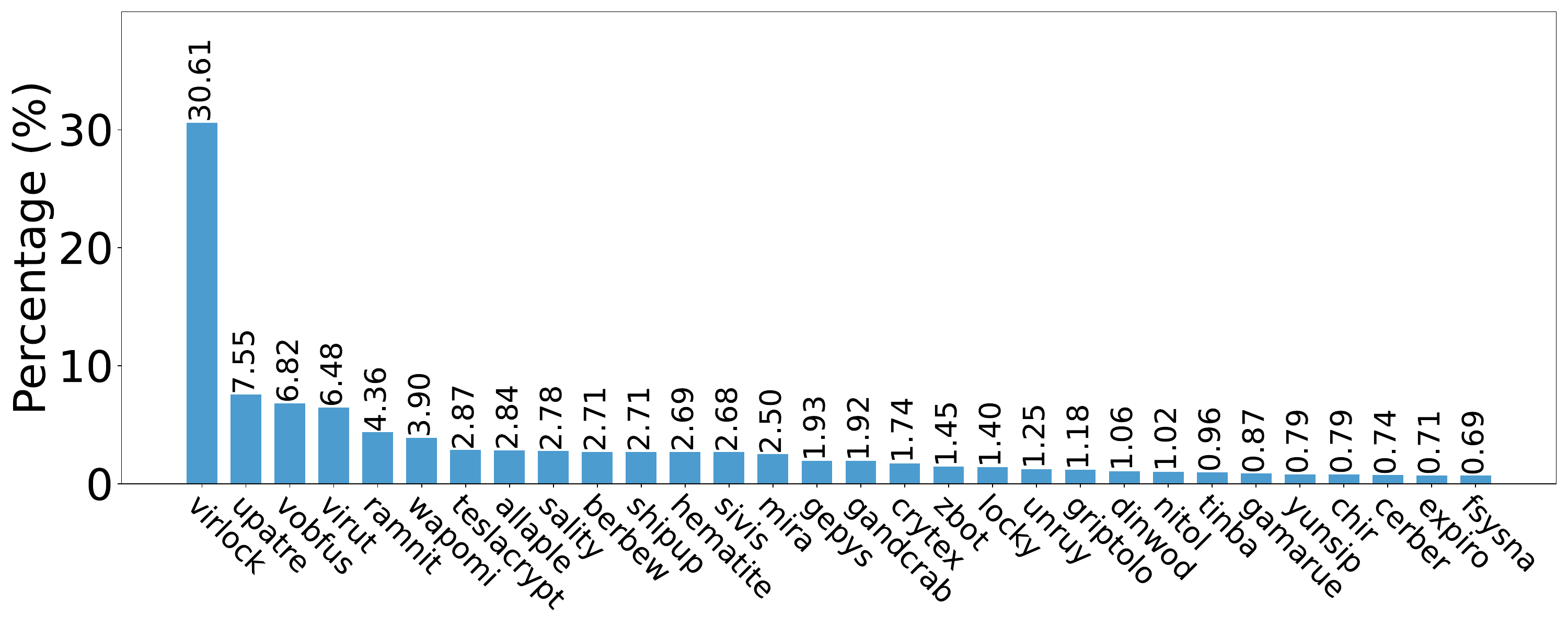}
    \caption{The distribution of top 30 malware families.}
    \label{figure:dataset}
\end{figure}

To fairly evaluate the effectiveness of learning-based malware detection systems in detecting previously unseen malware, we employ a time-based training/testing split.
In particular, for all malware, we take all malware samples before May 6, 2020, as the training set and equally divide the remaining malware into the validation and testing set. 
For all goodware, we randomly split them according to the percentages of the training/validation/testing sets of malware.
It is reported that there is a total of 848 different malware families, in which its training/validation/testing sets have 780/366/375 different malware families, correspondingly.
\Figref{figure:dataset} illustrates the distribution percentages of the top 30 malware families.

\section{More Evaluations}\label{appendix:section:more:evaluations}
The below includes four more evaluations as follows.

\textbf{(1) More evaluations in terms of size alteration ratio}.
In Table~\ref{tab:appendix:obfuscation:sizeAlterationRatio}, we measure the ``size alteration ratio (\%)'' of successfully evaded malware samples before and after the obfuscating process to observe the overhead introduced by the obfuscation tools.
It is clearly observed that the malware size alteration ratio after being obfuscated by UPX is approximately $3 \times$ times that of \mytool, while the size alteration ratios after being obfuscated by VMProtect and Enigma exceed $980 \times$ times and $330 \times$ times that of \mytool, respectively.

\begin{table}[htbp]
  \centering
  \small
  \setlength\tabcolsep{0.8pt}
  \caption{The ``size alteration ratio (\%)'' performance comparisons between \mytool and three baseline obfuscations.}
    \begin{tabular}{cccccccc}
    \toprule
    \multirow{2}[5]{*}{\tabincell{c}{Attacks}} & \multicolumn{2}{c}{MalGraph} & \multicolumn{2}{c}{Magic} & \multicolumn{2}{c}{MalConv}  & \multirow{2}[5]{*}{\tabincell{c}{Avg.}}\\
    \cmidrule{2-3} \cmidrule(l{1pt}r{1pt}){4-5} \cmidrule{6-7}   
    \multicolumn{1}{c}{} & \tabincell{c}{FPR\\=1\%} & \tabincell{c}{FPR\\=0.1\% }& \tabincell{c}{FPR\\1=\%} & \tabincell{c}{FPR\\=0.1\%}& \tabincell{c}{FPR\\=1\%} & \tabincell{c}{FPR\\=0.1\%} &  \multicolumn{1}{c}{}\\
    \midrule
    UPX         & -35.8     & -28.2     & -30.4   & -19.6    & -11.1   & -27.8    & 23.3   \\
    VMProtect   & —         & —         & +822.7  & +5343.8  & —       & —        & 7767.7 \\
    Enigma      & +1737.3   & +3850.5   & —       & +2089.7  & —       & +181.4   & 2613.1  \\
    \midrule
    \mytool    & +7.97     & +8.03     & +7.89   & +7.08    & +7.97   & +7.72    & 7.87   \\
    \bottomrule
    \end{tabular}%
  \label{tab:appendix:obfuscation:sizeAlterationRatio}%
\end{table}%

\textbf{(2) More evaluations of {\mytool} variants against anti-virus products}.
Table~\ref{tab:ablation:twokeycomponent:antivirus} summarizes the ASR performance of four \mytool variants against five anti-virus products.
In a similar vein, \mytool and \mytoolNOSpace(S) achieve the best attack performance than {\mytoolNOSpace$^{\dag}$} and {\mytoolNOSpace$^{\ddag}$} when evaluated by all employed real-world anti-virus products.
This observation reconfirms that, only by concurrently applying $\dag$ \underline{injecting semantic {\nop}s} and $\ddag$ \underline{redividing $\mathsf{call}$ instructions}, the attack effectiveness of \mytool can be maximized.

\begin{table}[htbp]
  \small
  \renewcommand\tabcolsep{2.8pt}
    \caption{The ASR performance (\%) of four \mytool variants against five anti-virus products.}
    \begin{tabular}{ccccccc}
    \toprule
     Attacks       &   McAfee  &   Comodo  &   Kaspersky   &   ClamAV      &   MS-ATP  \\
     \midrule
     {\mytoolNOSpace$^{\dag}$}   & 28.82 & 30.96 & 4.15 &  27.22  &  — \\
     {\mytoolNOSpace$^{\ddag}$}   & 44.03 & 33.41 & 6.05 &  29.75 &  — \\
    \midrule
    {\mytool}      &   48.81 &   36.00 &   11.29     &   31.94    &   {70.63}  \\
    {\mytoolNOSpace}(S)  &\bf   52.49 &\bf   36.36 &\bf   13.36     &\bf   32.33     &   \textbf{74.97}  \\
    \bottomrule
    \end{tabular}
    {\\\footnotesize{``—'' indicates that we cannot evaluate \mytool variants against MS-ATP as Microsoft has changed its authorization process since about 2024 and has not officially approved our application until now.}}
  \label{tab:ablation:twokeycomponent:antivirus}%
\end{table}

\begin{figure*}[htbp]
    \centering
    \begin{subfigure}{0.4\textwidth}
        \centering
        \includegraphics[width=\textwidth]{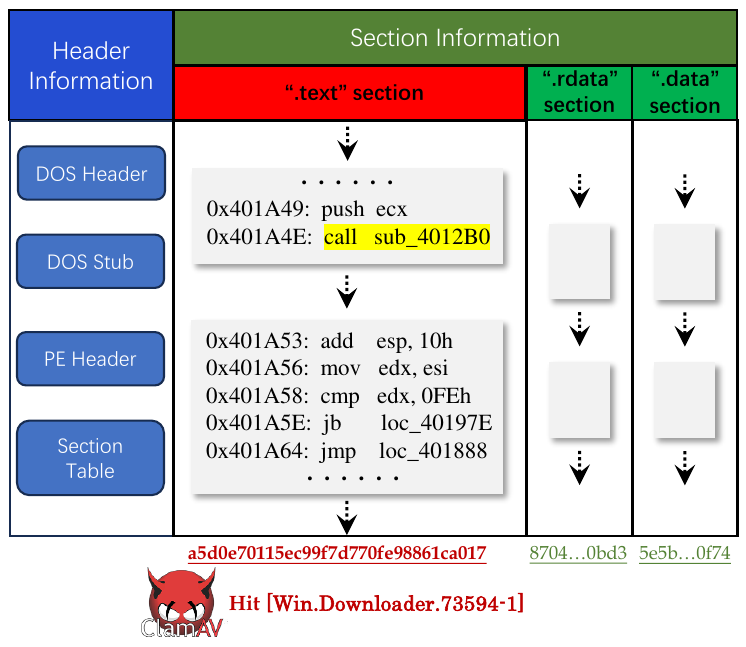}
        \caption{The ransomware is originally detected as malware.}
        \label{subfig:clamav:before}
    \end{subfigure}
    \hfill
    \begin{subfigure}{0.57\textwidth}
        \centering
        \includegraphics[width=\textwidth]{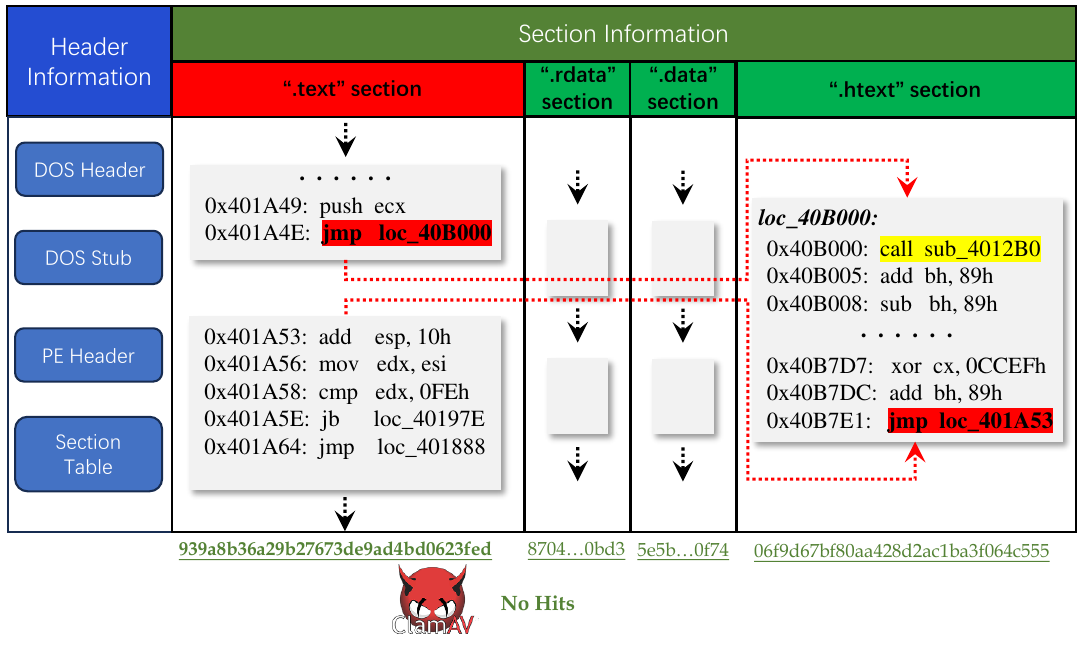}
        \caption{The corresponding adversarial ransomware is mistakenly identified as goodware.}
        \label{subfig:clamav:after}
    \end{subfigure}
    \caption{Illustration of a case study of how \mytool manipulates one ransomware to evade the detection of ClamAV.}
    \label{fig:clamav}
\end{figure*}

\textbf{(3) Impact of the size of semantic {\nop}s for {\mytool}}.
We also investigate the impact of the size of the injected semantic {\nop}s for {\mytool} against three target learning-based malware detection systems in the cases of FPR=1\%.
Recall from \Secref{subsec:method:reconstruction} that, the size of a newly injected section in Windows executables must be a multiple of the architecture's page size (\eg, 4KB for Intel x86).
To this end, we limit the size of semantic {\nop}s generated by {\mytool} to no more than $m \times 4KB$ (\ie, 1, 2, 3, 4, and 5) and present their attack performance against three target systems in Table~\ref{tab:different_settings_of_n}.
It is observed that, for both MalGraph and Magic, {\mytool} requires only one time the page size ($m=1$) to achieve a relatively high and stable ASR performance.
Differently, {\mytool} requires about three times the page size ($m=3$) to achieve a high and stable ASR performance against MalConv.
The main reason we conjecture is that MalConv purely relies on the raw bytes of malware, and MalGraph and Magic tend to rely on structural information like the function call relationship or control-flow information.
Thus, altering the detecting output of {MalConv} (\ie, evading its detection), requires {\mytool} to alter more raw bytes of the given malware, which is mainly accomplished by injecting semantic {\nop}s.
All the above indicates that \mytool can effectively evade learning-based malware detection with a small size of injected semantic {\nop}s.

\begin{table}[htb]
  \centering
  \small
  \renewcommand\tabcolsep{5pt}
  \caption{The ASR performance {\%} of {\mytool} against three target learning-based malware detection systems when the size of semantic {\nop}s to be injected is limited to $m \times 4KB$.}
    \begin{tabular}{cccccc}
    \toprule
    \tabincell{c}{Target Systems}  &   $m=1$    & $m=2$      & $m=3$       & $m=4$     & $m=5$     \\
    \midrule
    MalGraph                        &   96.31    &     97.49  &     97.60   &   97.54   &   97.61   \\
    Magic                           &   99.16    &     99.15  &     99.17   &   99.17   &   99.20   \\
    MalConv                         &   91.56    &     96.34  &     97.33   &   97.48   &   97.53   \\
    \bottomrule
    \end{tabular}%
  \label{tab:different_settings_of_n}%
\end{table}

\textbf{(4) Case study of how {\mytool} evade ClamAV}.
To unveil how \mytool evades real-world anti-virus products, we present a case study of how \mytool manipulates one wild malware to evade ClamAV.
We take ClamAV mainly because four of the five employed anti-virus products are proprietary and closed-source, and only ClamAV is open-sourced by Cisco~\cite{clamav_website}.
Notably, for ClamAV, an executable is reported as malware if any of its signatures match those in the ClamAV Virus Database (CVD), which is continuously updated with the latest malware samples.

To set up our case study, we randomly select one newly emerged malware variant from the WannaCry ransomware attack~\cite{wannacry}, termed as \textsc{WannaCry}\footnote{Its SHA245 hash is \hash{043bc5f8da479077084c4ec75e5c1182254366d-135373059906bb6fed0bf5148}}.
Subsequently, as illustrated in \Figref{subfig:clamav:before}, \textsc{WannaCry} is submitted to ClamAV and deterministically detected as malware because the hash signature of the ``.text'' section (\ie, ``\hash{a5d0e70115ec99f7d770fe98861ca017}'') exactly matches the signature in CVD.
Finally, as illustrated in \Figref{subfig:clamav:after}, after applying only a single {\call} transformation with {\mytool}, the hash signature of the ``.text'' section has been changed to ``\hash{939a8b36a29b27673de9ad4bd0623fed}'', which is no longer matched by any signature in CVD and is therefore mistakenly identified as goodware in the end.

\end{document}